# Calibrating parameters of crowd evacuation simulation at strategic, tactical and operational levels: Which one matters most?

Milad Haghani[a,*], Majid Sarvi[b]

a. School of Civil and Environmental Engineering, The University of New South Wales, UNSW Sydney, Australia
b. Department of Infrastructure Engineering, The University of Melbourne, VIC, Australia

milad.haghani@unsw.edu.au

## Abstract

*Background:* Simulating a process of crowd evacuation using an agent-based model requires the modeller to specify the value of a whole range of parameters each determining certain aspects of evacuees' behaviour. While potential sensitivity of simulation outputs to each single parameter is undoubted, one critical question is that 'to which class of parameters such numerical models are most sensitive when estimating evacuation times?' The question is of great significance given that available data and resources for calibration are limited and priorities often need to be made to calibrate model parameters. *Methods:* Here, we employ a simulation model of evacuation comprised of multiple integrated layers. After providing an overview of the calibration procedure for the key parameters of this model, we perform extensive sensitivity analyses by simulating evacuations in complex spaces with large crowds to answer the question raised above. *Findings:* Our analyses showed that, once parameters are contrasted on a consistent and comparable scale, estimates of simulated evacuation time are, by far margins, most sensitive to the value of mechanical movement parameters at 'locomotion' (or 'operational') layer, those that determine the discharge rate at bottlenecks. The next most critical set of parameters were those that determine the rate of 'direction choice adaptation' followed by those that determine the effect of 'inter-individual interactions' on direction choices. *Conclusions:* These results suggest that, in order to produce realistic evacuation time estimates using agent-based simulation models, the most critical aspect is, by far, perfecting the simulated flowrates at bottlenecks. *Implications:* This highlights the crucial need for furthering the accuracy of the existing empirical estimates for exit capacity; and developing more rigorous calibration methods as well as more nuanced settings for parameters that determine flowrates at bottlenecks. If a model does not produce bottleneck flowrates accurately enough, efforts to refine and sophisticate other aspects of simulations might be largely in vain.

# 1. Introduction

Similar to models of vehicular traffic, simulation models of crowd evacuations include a range of input parameters (Zeng et al., 2014). For microsimulation models, each of these parameters represent a certain aspect of the simulated agents' behaviour (Kretz et al., 2014; Wolinski et al., 2014). Unlike the case of vehicular traffic, however, the movement of pedestrians is less regulated and less constrained which adds more degrees of freedom to the movement and thereby more complexity to the simulation procedure. In a crowd evacuation scenario, pedestrians can choose their own speed of movement within the physical boundaries of the space and their movement is not constrained to a pre-defined physical network. They can overtake others (Zhang et al., 2018) and reverse/adjust their decisions (Gwynne et al., 2000) and movement trajectories (Corbetta et al., 2015). Their actions and their perceived 'best strategy' are influenced by observing the decisions of others (Kinateder et al., 2018; Kinateder et al., 2014a; Kinateder et al., 2014b) which, in addition to the heightened state of emotion caused by urgency (Moussaïd et al., 2016; Muir et al., 1996; Shipman and Majumdar, 2018), makes the behavioural response complicated thereby making the simulation a challenging task.

Studies have time and again shown how the prediction outcomes of these models depend heavily on the value of their input calibration parameters (Crociani et al., 2018; Haghani et al., 2018a; Hollander and Liu, 2008). Great efforts have been made in the past in the scientific literature to develop mathematically sophisticated models of simulation that can potentially represent a large variety of behavioural phenomena (Teknomo, 2006). But not as much attention has been paid to the problem of fine-tuning the parameters of such models (Dias et al., 2018; Dias and Lovreglio, 2018; Ko et al., 2013; Kretz et al., 2018; Li et al., 2015; Werberich et al., 2015a). Due to the lack of benchmark empirical data, in many cases, these calibration parameters have been set arbitrarily purely based on best suggestions of the modellers (Helbing et al., 2000). One may also argue that the problem of 'model validation' is just as much an overlooked issue (Berrou et al., 2005; Gwynne et al., 2005; Lovreglio et al., 2014b), which is true. Although, that problem is beyond the scope of the current study.

Within a given structure of simulation modelling, one would be able to produce vastly different patterns of movement and thereby vastly different predictions of evacuation time depending how the parameters of the model are set. Due to the variety of the existing simulation models, no formal or universal guideline has thus far been established for the calibration of simulation models (Hollander and Liu, 2008). Often, the procedure for parameter calibration is known and given but the required data may not exist in adequate quality and quantity. As a result of that, "in practice, simulation model–based analyses have often been conducted under default parameter values or best guessed values" (Park and Schneeberger, 2003). According to these authors, "This is mainly due to either difficulties in field data collection or lack of a readily available procedure for simulation model calibration and validation".

The general behaviour of an evacuee in real-life is comprised of multiple aspects of decision-making (Gwynne and Hunt, 2018), and so should be the procedure for simulating the behaviour of simulated evacuees within agent-based microscopic models (Asano et al., 2010; Zeng et al., 2017). As a result, a comprehensive representation of evacuees' behaviour in a simulated modelling framework requires a large number of parameters. These parameters each bear physical/mechanical or behavioural interpretations and are all crucial for determining how accurately the model predicts (Li et al., 2015; Lovreglio et al., 2015; Werberich et al., 2015b). However, there is no reason to believe that they all share the same level of significance in terms of influencing the overall accuracy of the model. It is not inconceivable that some parameters may play a more critical role than others in terms of determining how accurately the model predicts. Hence, a legitimate question that one might ask in this regard would concern the identification of those most critical parameters. The answer to this question is of great significance and has major implications for modelling practices in that it gives the modeller guides as to which kinds of parameters require more attention (e.g. more rigorous calibration procedure, more or better-quality data and more validation testing). Fine-tuning calibration parameters has always been challenging for variety of reasons ([Kretz et al., 2018](#)) including unavailability of the required data, but knowing which parameters need prioritisation can be of great value to inform the modeller where the most attention and resources need to be deployed. (e.g. in terms of provision of field ([Robin et al., 2009](#)) or experimental ([Daamen and Hoogendoorn, 2012a](#)) data).

This study aims to address the abovementioned question in particular. In this work, we employ a relatively comprehensive tool of simulation modelling that we have developed which is composed of all major layers of behaviour (or modules of modelling) relevant to crowd evacuation. Most layers of this simulation tool are econometric (choice or duration) models and nearly all layers are fully parameterised. Most critical parameters of the model have been directly calibrated based on disaggregate experimental data. In this work, we first provide a brief overview of the calibration procedure of this model and will subsequently focus on the main question that drives this study: which set of parameters are most critical in determining the simulation outputs? To explore this question, we design an extensive set of sensitivity analyses of these parameters around their calibrated values in a way that makes the relative comparisons between them possible. In consideration of the fact that outcomes of such kind of analyses may be case specific and dependant on the particular features of the simulated setup, we examine a large variety of relatively complex crowd evacuation setups. The unilateral sensitivity of the predictions to each of these parameters when they are stretched enough, and particularly in simple geometries, have been established in earlier work and are not doubted ([Haghani and Sarvi, 2019a](#); [Haghani et al., 2018a](#); [Haghani et al., 2018b](#)). What particularly concerns us here is what happens if these parameters are contrasted on a comparable scale and what happens if the scenarios are not simple rooms. We intend to examine scenarios that show more resemblance to real-world spaces and those that are complex enough to require all available layers of the model to remain active and essential for the simulation. Such analyses will give us an indication of the kind of parameters that have the highest priority to be calibrated in cases of evacuation simulation in complex spaces. Also, by association, such findings will provide a range of behavioural insight as secondary outcomes. They will, for example, provide indications about important question such as the followings. In order for the process of a crowd evacuation to be accelerated, what aspects of evacuees' behaviour can be modified, and in what direction. What the priorities would be in terms of influencing evacuees' behaviour? Whish aspect of their behaviour will have the largest impact on accelerating the discharge process if modified? This has great implications for response agencies and evacuation managers whose aim is to optimise evacuation processes and increase the chance of survival in crises.

## 2. Simulation model structure
### 2.1. Overall structure

The simulation model that we employ for the numerical analyses reported in this work is comprised of six major modelling layers. These include reaction time module, room-sequence-choice module, exit-choice module, exit-choice changing module, local pathfinding and step-taking module. Consistent with a number of earlier studies, we adopt the three-level categorisation of 'strategic', 'tactical' and 'operational' in order to refer to the various layers of this model (Haghani and Sarvi, 2018a). Any layer of the model that generates a decision prior to the actual movement of the simulated agents is labelled a 'strategic' layer (this includes the reaction time and room-sequence choice). The exit-choice model and the model that allows those choices to be revisited and possibly updated are labelled 'tactical' layers of the model. The local pathfinding and step-taking models are 'operational' layers of the model and are the modelling layers that mechanically navigate and move each agent towards their chosen target in accordance with all their major decisions have been previously simulated by the two other layers. Figure 1 provides a summary of this modelling structure and their associated parameters. Tables 1 and 2 respectively details the definition of the parameters and the variables included in these modelling layers.

We acknowledge that such categorisation can be rather arbitrary and may depend on the structure of each simulation framework. Currently, there is no universal standard for such categorisation and there may not be a need for one given that such categorisation does not serve any purpose other than facilitating the communication of the model structure to readers.

The strategic, tactical and operational decisions are generated in sequence and individually for each simulated agent. Each layer of the model provides part of the necessary input for the layers that comes immediately after. The strategic and tactical modules are all econometric (duration or choice) type of models. The local pathfinding module is algorithmic, and the step-taking module is simply a social-force model. Only a subset of the parameters has been directly calibrated based on empirical observations (those that were assumed to be most critical and those for which experimental data could be collected). The asterisk sign in Table 1 specifies these calibrated parameters. For the rest of the parameters, reasonable default values have been suggested based on trial and error testings during the process of development and implementation. Only parameters that have been directly calibrated are the subject of sensitivity analyses in this work.

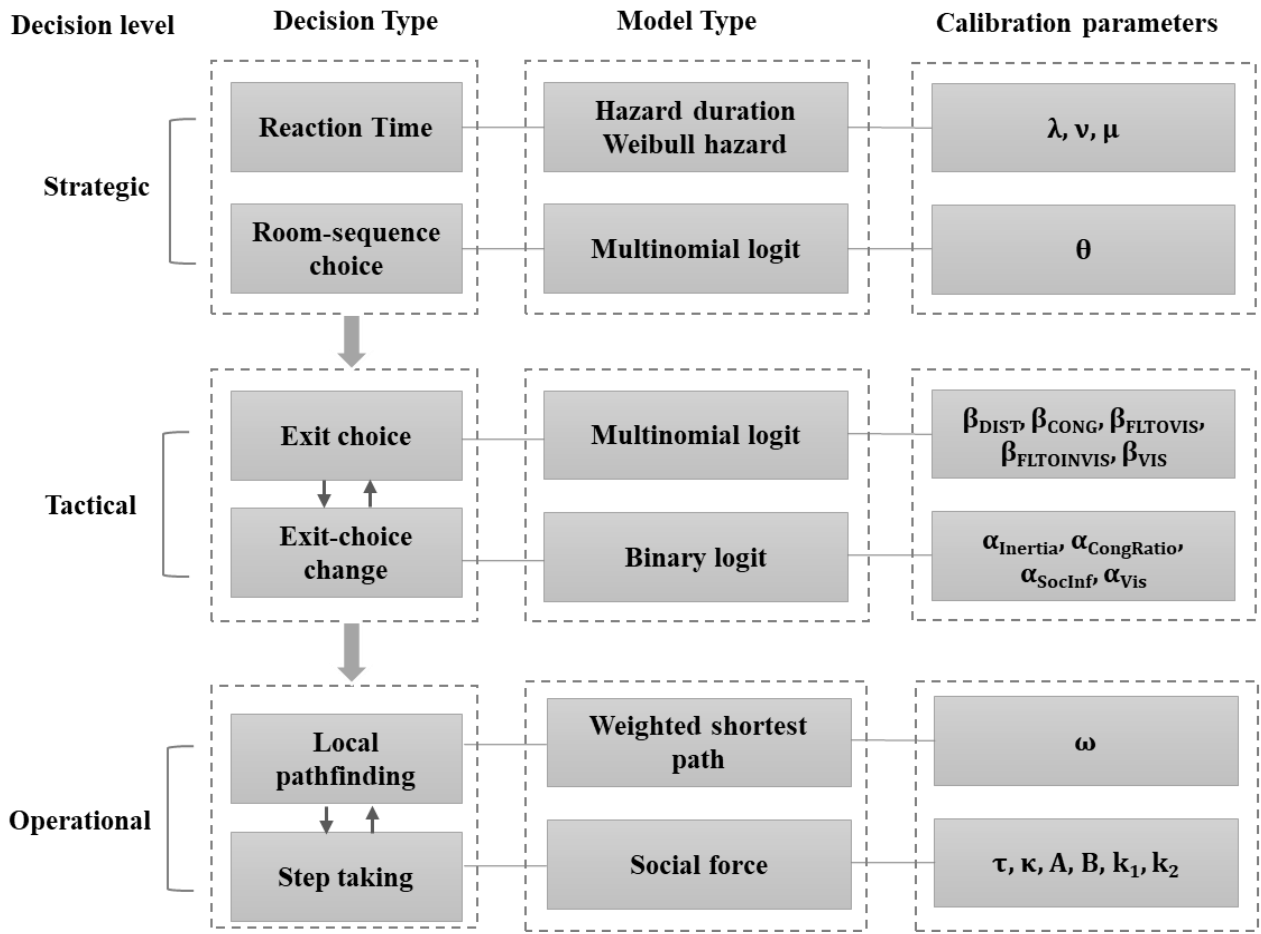

**Figure 1** The overall structure of the computational simulation modelling tool employed in this work. Illustrating various behavioural layers of the model, the nature (or type) of the model at each layer and the parameters involved in each layer.

**Table 1** The list of the parameters of the simulation model, their descriptions and the calibrated (or suggested) values. The asterisk indicates the parameters whose values have been calibrated empirically and are the subject of the sensitivity analyses.

| Parameter | Description | Calibrated value |
|---|---|---|
| $\lambda$ | Parameter related to the mean and variance of the reaction times. Default value is positive. Greater values mean smaller reaction times as well as less variability in reaction times across simulated occupants. | 1.523* |
| $\nu$ | Parameter related to the reaction times. Default value is positive. Not subject of the sensitivity analyses. | 2.511 |
| $\mu$ | Parameter that determines the effect of distance to nearest exit (at the onset of evacuation) on reaction time. Default value is negative. Greater absolute values mean that the further the agent is to exits, the longer it takes for it to initiate movement. | -0.305* |
| $\theta$ | Parameter related to the room sequence choice reflecting the effect of distance to the final gates on room sequence choice. Default value is positive. Greater values mean that the agent is more likely to choose the room sequence that offers the minimum distance to a final exit. | 0.500 |
| $\beta_{DIST}$ | Parameter related to the exit choice reflecting the effect of the distance to exits on exit choice. Default value is negative. Greater absolute values mean that the agent is more likely to choose the nearest distance. | -0.278* |
| $\beta_{CONG}$ | Parameter related to the exit choice reflecting the effect of the queue size (congestion) on exit choice. Default value is negative. Greater absolute values mean that the agent is more likely to choose the least congested exit. | -0.158* |
| $\beta_{FLTOVIS}$ | Parameter related to the exit choice reflecting the effect of the flow size on exit choice (for the exits that are visible). Default value is negative. Greater absolute values mean that the agent is more likely to avoid the visible exits to which heavy flows are moving. | -0.042* |
| $\beta_{FLTOINVIS}$ | Parameter related to the exit choice reflecting the effect of the flow size on exit choice (for the exits that are invisible). Default value is positive. Greater values mean that the agent is more likely to follow the invisible exits to which heavy flows are moving. | 0.013* |
| $\beta_{VIS}$ | Parameter related to the exit choice reflecting the effect of the exit visibility on exit choice. Default value is positive. Greater values mean that the agent is more likely to choose the exits that are visible. | 0.802* |
| $\alpha_{Inertia}$ | Parameter related to the exit-choice-changing behaviour reflecting the tendency of the agent to maintain the initial choice of exit. Default value is negative. Greater absolute values mean that the agent is more likely to choose to not to change the previous choice of exit when given the opportunity to revise that decision. | -7.234* |
| $\alpha_{CongRatio}$ | Parameter related to the exit-choice-changing behaviour reflecting the relative queue size at the chosen exit compared to that of the least congested exit in that room. Default value is positive. Greater values mean that the agent is more likely to choose to change the previous choice of exit when the queue at that exit is relatively large. | 3.119* |
| $\alpha_{SocInf}$ | Parameter related to the exit-choice-changing behaviour reflecting the social (or peer) influence. Default value is positive. Greater values mean that the agent is more likely to choose to change the previous choice of | 0.577* |
| | exit when other agents have made the same change during earlier moments. | |
| $\alpha_{Vis}$ | Parameter related to the exit-choice-changing behaviour reflecting the exit visibility effect. Default value is positive. Greater values mean that the agent is more likely to choose to change the previous choice of exit when a less congested exit in that room is within its visual field. | 1.876* |
| $\omega$ | Penalty weight for the weighted shortest path algorithm. Greater values mean that the agent will avoid the areas occupied by other agents and takes longer detours to reach the chosen exit. | 4500 |
| $\tau$ | Relaxation parameter of the social-force model. Default value is positive. Greater values mean that the driving force is weaker, and that the bottleneck flow rate is slower. | 0.120* |
| $\kappa$ | Friction force parameter of the social force model. Default is positive. Greater values mean that the friction force between the agent and other agents or walls that are in physical contact with the agent is greater, and that the bottleneck flow rate is slower. | 5500* |
| $A$ | Social force parameter related to interindividual and wall forces when there is no physical contact. Default is positive. Greater values mean that the agent perceives greater repulsion forces from other agents or the walls when there is no contact with them. | 2000 |
| $B$ | Social force parameter related to interindividual and wall forces. Default is positive. Greater values mean that the agent perceives smaller repulsion forces from other agents or the walls when there is no contact with them. | 0.080 |
| $k$ | Social force parameter related to interindividual and wall forces. It determines the obstruction effects in cases of physical interactions | 120000 |

**Table 2** The list of the variables and their description in the simulation model

| Variable | Description |
|---|---|
| $Min\_Exit\_DIST_i$ | Euclidian distance of agent $i$ to the nearest exit in the current room at the onset of the evacuation |
| $Room\_Seq\_DIST_{ir}$ | The distance from the position of agent $i$, to the nearest final gate through room sequence $r$, at the onset of the evacuation |
| $DIST_{ie}$ | The Euclidian distance from the position of agent $i$, to exit $e$ |
| $CONG_{ie}$ | The queue size (congestion) at exit $e$ observed by agent $i$ |
| $FLOW_{ie}$ | The flow size moving to exit $e$ observed by agent $i$ |
| $VIS_{ie}$ | The visibility status of exit $e$ according to agent $i$, equals 1 if visible and 0 otherwise |
| $Min\_CONG_i$ | The queue size at the least congested exit in the current room of agent $i$ |
| $CONG_{ie^*}$ | The queue size at the currently chosen exit $e^*$ by agent $i$ |
| $Vis_i$ | The congestion-visibility status for agent $i$, equals 1 if there is a visible AND less congested exit available at the current room for agent $i$, and 0 otherwise |
| $SocInf_i$ | The social influence on exit-choice change for agent $i$. The number of other agents changing their decisions from $e^*$ to another exit within the last $t=2$ seconds of the simulated time |
| $m_i$ | The body mass of agent $i$ |
| $\vec{r}_i$ | The body ratio of agent $i$ |
| $v_i^d$ | The desired velocity of agent $i$ |
| $\vec{e}_i^d$ | The unit vector of desired direction for agent $i$ |
| $d_{ij}$ | The Euclidian distance between agents $i$ and $j$ |
| $d_{iw}$ | The Euclidian distance between agent $i$ and wall $w$ |

## 2.2. Reaction time

The reaction times of the simulated agents are generated using a Cox-proportional Weibull Hazard Duration model. For each simulated agent $i$, the minimum distance to exits in the initial room where the agent is located, at the onset of the evacuation, is measured ($Min\_Exit\_DIST_i$). This is simply calculated by measuring the Euclidian distance between the spatial coordinates of the agent and those of the centre point of each exit in the current room. The amount of time difference between the onset of the simulated evacuation and the moment the agent initiates a movement, known as the agent's reaction time $T_i$, is determined by the following equation.

$$T_i = \left(-\frac{\ln u}{\lambda \exp(\mu \times (Min\_Exit\_DIST_i))}\right)^{1/\nu} \quad (1)$$

Where:

$$u \sim U(0,1) \quad (2)$$

## 2.3. Room-sequence choice

The room-sequence choice is a multinomial logit model. At the onset of evacuation (before any agent initiates its movement), the set of room sequences that can lead an agent from the current room of agent $i$ to a final exit (i.e. hypothetical safety) is enumerated. We label this set by $R_i$. For each agent $i$ and for each of these room sequences like $r$, the minimum distance between the coordinates of agent $i$ and a final exit is calculated and labelled as $Room\_Seq\_DIST_{ir}$. To calculate these variables, each room sequence in $R_i$ is considered, and within that room sequence, all possible exit sequences are enumerated. The one that generates the smallest distance is outputted as $Room\_Seq\_DIST_{ir}$. It should be noted that this is only a rough measurement in order to generate a reasonable room sequence for the agent. Through the exit-choice changing module, the agent will later during the movement process be able to modify this sequence in a dynamic way. The probability of agent $i$ choosing room sequence $r$ is given by the following equation. The chosen room sequence is simulated for each agent according to the sets of room-sequence probabilities for that agent.

$$P_{ir}^{room} = \frac{\exp(-\theta \times Room\_Seq\_DIST_{ir})}{\sum_{s \in R_i} \exp(-\theta \times Room\_Seq\_DIST_{is})} \quad (3)$$

## 2.4. Exit choice

Once the choice of room sequence is simulated for an agent $i$, the information is fed to the exit-choice module as an input. This module first generates the set of all exits in the current room (where the agent is located), those that are feasible according to the previous choice of room sequence. These exits constitute the set of feasible exits for agent $i$, $E_i$. At the simulated moment when the choice of exit is generated for agent $i$, the attributes of all exits are calculated for agent $i$. This includes $DIST_{ie}$ (distance to exit $e$), $CONG_{ie}$ (congestion or queue size at exit $e$), $FLOW_{ie}$ (flow size towards exit $e$) and $VIS_{ie}$ (dummy 0-1 variable that equals 1 if exit $e$ is visible to agent $i$). Then interaction of the $FLOW$ and $VIS$ are calculated through simple transformations to generate $FLTOVIS$ and $FLTOINVIS$ attributes: $FLTOVIS_{ie}=FLOW_{ie} \times VIS_{ie}$ and $FLTOINVIS_{ie}=FLOW_{ie} \times (1-VIS_{ie})$. Having calculated the attributes of all exits in the choice set of agent $i$, the probability of choosing exit $e$ is calculated using the following equations. The chosen exit is simulated for each agent according to the sets of exit probabilities for that agent.

$$P_{ie}^{exit} = \frac{exp\,(V_{ie})}{\sum_{f \in E_i} exp\,(V_{if})} \quad (4)$$

Where:

$$V_{ie} = \beta_{DIST}(DIST_{ie}) + \beta_{CONG}(CONG_{ie}) + \beta_{FLTOVIS}(FLTOVIS_{ie}) + \beta_{FLTOINVIS}(FLTOINVIS_{ie}) + \beta_{VIS}(VIS_{ie}) \quad (5)$$

*2.5. Exit-choice changing*

The initial exit choice generated for each agent *i*, is allowed to be revisited at a certain time frequency. Agent *i* is given the possibility to revisit its initial choice of exit and it is given a binary choice between two alternatives: 'change' or 'not-change'. In order to simulate the probabilities for these two alternatives, a number of variables (associated with agent *i*) are calculated at the moment of decision revisit. This includes $Min\_CONG_i$ (the smallest size of queue (smallest congestion) among the all the exits in the current room where *i* is located), $CONG_{ie*}$ (the queue size at the currently chosen exit for agent i, where $e^*$ indicates the chosen exit), $Vis_i$ (a dummy 0-1 variable that equals 1, if there is a visible exit with smaller queue size than that of the currently chosen exit for *i*), $SocInf_i$ (the number of agents (other than *i*) that have changed their choice of exit from the one currently chosen by *i* to another exit within the last *t=2* seconds of the simulated time). The probability of agent *i* changing its initial exit choice is calculated using the following equations. The choice of whether to change or not change is simulated according to these probabilities. If the simulated decision is to change the initial choice, then a new choice set is generated for agent *i*, and that choice set is fed back to the exit-choice module to generate a new choice of exit for *i*. This new choice set excludes the one exit that is chosen currently and also the ones through which the agent has entered the current room. It includes every other exit in the current room that can lead to any final exit. This implicitly allows the initial choice of room-sequence to also be revisited at that instance.

$$P_{change_i} = \frac{1}{1 + e^{-V_{change_i}}} \quad (6)$$

Where:

$$V_{change_i} = \alpha_{Inertia} + \alpha_{CongRatio} \underbrace{\left(\frac{CONG_{ie*} - (Min\_CONG_i)}{CONG_i}\right)}_{=CongRatio_i} + \alpha_{Vis}(Vis_i) + \alpha_{SocInf}(SocInf_i) \quad (7)$$

*2.6. Local pathfinding*

Once the choice of exit (an initial choice or an updated one) is simulated, the information is passed on to the subsequent layer of the model, the local pathfinding. This module is algorithmic and not based on a model like the previous layers and is borrowed from a popular algorithm in computer science and particularly computer gaming. The A*-pathfinding algorithm, as an extension of the Dijkstra's shortest path algorithm, has been adopted for this module (Cui and Shi, 2011). The algorithm basically generates the shortest weighted path between the current coordinates of the simulated agent (the origin node) and the mid-point coordinates of the chosen exit in the current room

(the destination node). The shortest weighted path is generated based on a weighted grid graph overlaid on the movement space. The weighted grid graph allows the simulated agent to circumvent congested areas by penalising the links proportional to the number of other agents in their vicinity. The closest node to each agent in the room is identified and all links leading to that node are penalised by a factor of $\omega$. These penalties are cumulative. The presence of the obstacles penalises the nodes in the same way except that in this case, the magnitude of the penalty is infinite so that the agent completely avoids the obstacles. The A* shortest path for each agent is updated at a certain frequency. Once the path is generated for an agent, the vector that connects the current coordinates to the coordinates of the next node on the path is calculated and fed to a social-force model (Helbing et al., 2000) as the vector of 'desired direction'.

## 2.7. Step-taking

For each agent $i$, each step motion is determined by the net outcome ($\vec{F}_i^{net}$) of the desired force of pedestrian $i$, $\vec{F}_i^d$, the social forces from other pedestrians $j \neq i$, $\vec{F}_{ij}$; and the forces from walls ($w$), $\vec{F}_{iw}$. These forces are formulated through the following equations according to Helbing et al. (2000). Definitions of the parameters and variables in these equations can be found in Tables 1 and 2 respectively. The function $G(r_i + r_j - d_{ij})$ is an identity function when its argument is positive, and otherwise, returns the value of 0. Also, $\tau_i, A_i, B_i, k$ and $\kappa$ are calibration parameters of this model.

$$\vec{F}_i^{net} = m_i \frac{d\vec{v}_i}{dt} = \vec{F}_i^d + \vec{F}_{ij} + \vec{F}_{iw} \qquad (8)$$

$$\vec{F}_i^d = m_i \frac{v_i^d \vec{e}_i^d - \vec{v}_i}{\tau_i} \qquad (9)$$

$$\vec{F}_{ij} = \left(A_i e^{(r_i+r_j-d_{ij})/B_i} + k\, G(r_i + r_j - d_{ij})\right)\vec{n}_{ij} + \kappa G(r_i + r_j - d_{ij})\Delta v_{ji}^t \vec{t}_{ij} \qquad (10)$$

$$\vec{F}_{iw} = \left(A_i e^{(r_i-d_{iw})/B_i} + k\, G(r_i - d_{iw})\right)\vec{n}_{iw} - \kappa G(r_i - d_{iw})(\vec{v}_i \cdot \vec{t}_{iw})\vec{t}_{iw} \qquad (11)$$

$$\vec{n}_{ij} = \frac{1}{d_{ij}}(\vec{r}_i - \vec{r}_j) = (n_{ij}^1, n_{ij}^2) \qquad (12)$$

$$\vec{t}_{ij} = (-n_{ij}^2, n_{ij}^1) \qquad (13)$$

$$\Delta v_{ji}^t = (\vec{v}_j - \vec{v}_i) \cdot \vec{t}_{ij} \qquad (14)$$

## 3. Overview of the calibration

Each layer of the model outlined in the previous section was calibrated directly through a certain set of experimental observations. This is with the exception of the room-sequence choice module and the local pathfinding module. The empirical observations were collected through series of simulated evacuation experiments that the authors have carried out since 2015 (Figures 2-5). With the exception of the bottleneck experiment (Figure 5), each experiment was designed for analysing and extracting disaggregate data for certain aspect(s) of decision making. The data for those experiments were obtained through individual-by-individual image processing and choice data extraction. We label these four experiments, illustrated in Figures 1-4 respectively as Experiment I, II, III and IV. Overall details for these experiments have been provided in Table 3 and full details can be accessed in previous individual publications ([Haghani and Sarvi, 2018b](); [Haghani et al., 2019a](), [b]()).

Experiments I and II were primarily designed to generate exit-choice observations. Experiment III was primarily designed to generate reaction time and exit-choice change observations. However, both Experiments I and II were subsequently re-analysed for observations of exit-choice change and contributed observations to that dataset. Experiments II and III scenarios were carried out under two distinct contextual conditions: low urgency and high urgency. Here, we only focus on simulation of a high-urgency evacuation and therefore, our calibration estimates are reported consistent with that. Therefore, the segment of observations that were derived from low-urgency scenarios in each of the experiments II and III were disregarded for this analysis. Similarly, Experiment IV scenarios were conducted under three levels of competitiveness and here we only use the part of the observations that is derived from the highest level of competitiveness. Also, this experiment was only analysed at the aggregate level and based on the single measure of 'total evacuation time'.

For the decision-making experiments (experiments I, II and III) each scenario is unique and is a combination of a certain set of control factors (such as the number of exits, the widths of exits, the positions of open exits, the number of participants and the positioning of barricades). Since the experiments were designed (primarily) for individual-level data extraction, they were exempt from requiring repetition of individual scenarios. What is critical for such design is to generate large degrees of variability for the individual-level observations, a design attribute that was achieved through changing the levels of those control factors. This ensures that the data overall contains a large variety of choice observations (in terms of the attribute levels and the size of the choice set) required for efficient model estimation.

In experiments I and II, subjects would wait at certain holding areas (one holding area for Experiment I and two for Experiment II) and entre the evacuation room where they have multiple exit choice options. During the data extraction process, we would single out the trajectory of each participant one by one, analyse their movement and identify the moment where their trajectory, body movement, head orientation and step size indicate that a firm exit decision has been made. At that moment during the movement of the subject, we would freeze the scene and extract the information relevant to their choice including the choice itself, the choice set and the attribute levels (or variables) (as specified by Eq. 5). This set of information constitutes one choice observation. Should the participant display more than one clear choice observation (which happened occasionally during the experiments), then we would extract a secondary observation for that subject as well.

For the subjects that show multiple exit decisions (i.e. those who make a clear initial decision and advance towards it and then subsequently, change their decision to another exit), we also extract exit-choice change observations. For these subjects the moment of the secondary exit choice coincides with the moment of the exit-choice change. We would identify the moment at which the subject abandons the previously-chosen exit and make advancement towards another alternative. At this moment, as pointed out earlier, a secondary exit choice observation is extracted. While still freezing the scene at that moment (which we refer to as the moment of the 'decision change'), we also extract an 'exit-choice change' observation. This observation is a binary choice, and for that, we record the following set of information: the choice (which is always 'change' for the subject at hand), the choice set (which is fixed consisting of the 'change' and 'no change' alternatives), and the attribute levels (as specified by Eq. 7). While at that moment, one auxiliary 'no-change' observation is also extracted from any other exit in that space (other than the one that the subject has abandoned) so that the choice contains both change and no-change observations. This process was first applied to the observations of Experiment III, but then was subsequently also applied to Experiments I and II that were re-analysed for this purpose.

For the extraction of the reaction time observations in Experiment III, we also singled out each individual's trajectory while playing the scene in PeTrack (Boltes and Seyfried, 2013). For each subject, we identify the moment at which the subject initiates his/her (decisive) movement towards an exit. At this point, we would freeze the scene and record the frame number associated with that moment. By comparing that frame number to the frame number associated with the onset of the evacuation (i.e. the moment we produced the evacuation signal), we would calculate the difference and covert it to a reaction-time observation in terms of seconds. At that moment in time, we would also record the spatial coordinates of the subject at hand. By comparing those coordinates with the (mid-point) coordinates of the open exits (in that scenario), the distance to each exit is calculated. Through this calculation, the distance to the nearest exit is recorded for that reaction-time observation, consistent with our formulation in Eq. 1.

Through these individual-level data extraction processes, 3685 observations of exit choice (3015 from Experiment I, and 670 from experiment II), 618 observations of exit-choice change (125 from Experiment I, 85 From Experiment II and 408 from Experiment III) and 851 observations of reaction time (all from Experiment III) were generated in total. Each of these three sets of observations were used to estimate their respective models using maximum likelihood estimation method (Haghani et al., 2019a; Haghani et al., 2016). All estimation calculations were performed using NLogit software (Greene, 2007).

The two coefficients of the social force model, $\tau$ and $\kappa$ were calibrated through seven individual observations of the total evacuation time generated from the bottleneck experiment (Experiment IV, Figure 5) (Haghani et al., 2019c). The calibration process was based basically on minimising the error between the simulated and observed total evacuation times at each different exit width. In this experiment, the exit width ranges from 60cm to 120cm at the increments of 10cm. Unlike the decision-making modules, this layer was not directly calibrated from disaggregate data. Rather, it was calibrated through comparisons between the simulated and observed measurement of a certain aggregate metric, total evacuation time. Full details of this calibration can be accessed in (Haghani and Sarvi, 2019b). The overall process involved selecting a certain range for both $\tau$ and $\kappa$ values, considering all possible combinations between these two parameters within those ranges and simulating each experimental scenario (for 50 times) for each combination. The combinations that resulted in the least amount of relative error percentage (i.e. percentage difference between observed and simulated evacuation time associated with that exit width) were 'shortlisted' and the combinations that resided at the intersect of all 'shortlists' were selected as the calibrated values.

Figure 6 provides a visualisation illustration of the outcomes of this analysis in a heatmap-type plots. Each graph is associated with a certain width of exit, and the horizontal and vertical axes represent the parameter values for κ and τ. The simulation error associated with each parameter value has been colour-coded, with brighter colours representing lesser error. As a result, the areas of each plot that are represented by white colour are associated with the τ-κ combinations that produce the least amount of simulation error for that exit with (i.e. produce the best match with the experimental observations). Only a few combinations resided at the intersect of those white regions for all seven widths. The chosen values reported in Table 1 is among those, though those combinations were not unique. There were other combinations that were almost equally good in minimising the simulation error.

**Table 3** The list of the experiments whose observations were used for the calibration of the numerical simulation model used in this study.

| Experiment No. | Year | No. of Participants | No. of Scenarios | Relevance to calibration |
|---|---|---|---|---|
| I | 2015 | 142 | 29 | Exit choice<br>Exit choice change |
| II | 2017 | 114 | 24 | Exit choice<br>Exit choice change |
| III | 2017 | 146 | 24 | Reaction time<br>Exit choice change |
| IV | 2017 | 114 | 21 | Social force (at bottlenecks) |

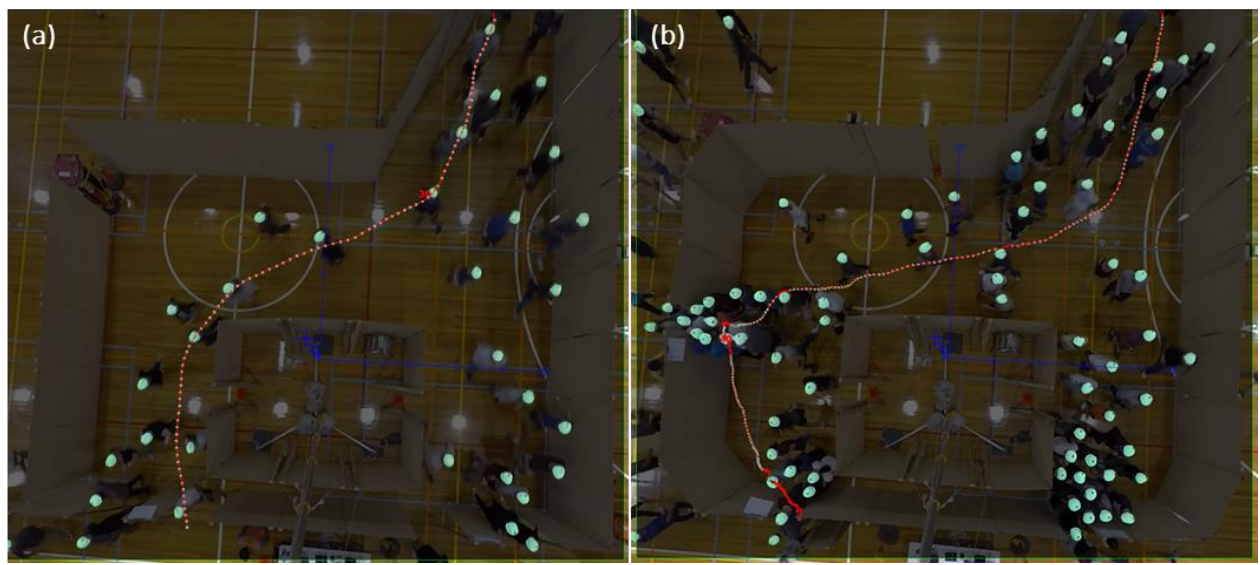

**Figure 2** Still images from the data extraction process of a sample of scenarios in Experiment I. Subfigure (a) shows an exit-choice observation extraction example and subfigure (b) shows an example of data extraction for exit choice adaptation. Both figures only illustrate the moment of observation extraction. At such moments, the set of information regarding the subject's decision is recorded including the choice, choice set and attribute levels of the alternatives.

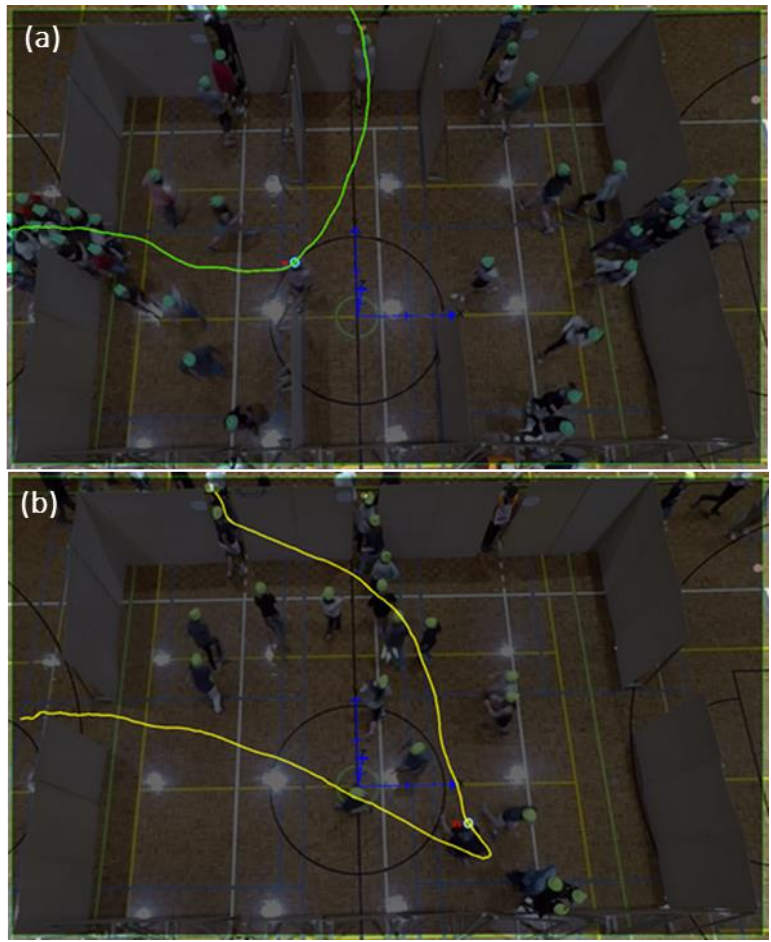

**Figure 3** Still images from the data extraction process of a sample of scenarios in Experiment II. Subfigure (a) shows an exit-choice observation extraction example and subfigure (b) shows an example of data extraction for exit choice adaptation. Both figures only illustrate the moment of observation extraction. At such moments, the set of information regarding the subject's decision is recorded including the choice, choice set and attribute levels of the alternatives.

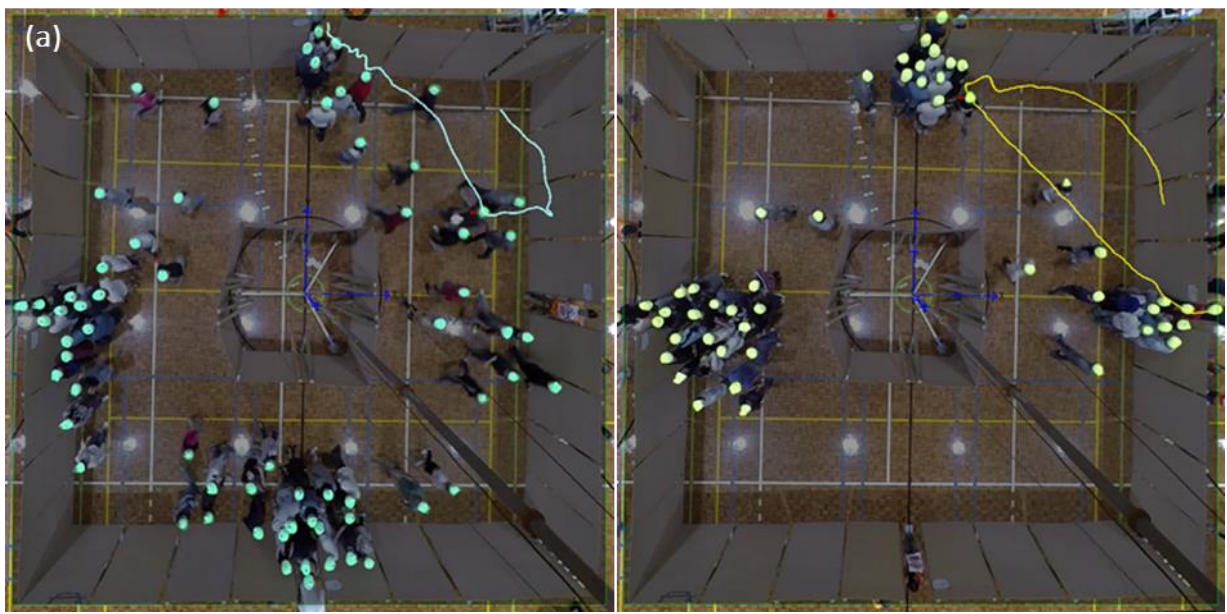

**Figure 4** Still images from the data extraction process of a sample of scenarios in Experiment III. Subfigure (a) shows a reaction-time observation extraction example and subfigure (b) shows an example of data extraction for exit choice adaptation.

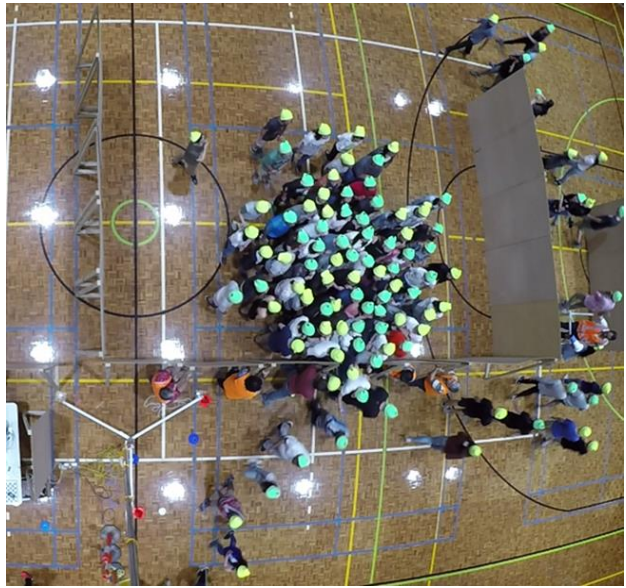

**Figure 5** Still image from the setup of the Experiment IV, the bottleneck experiment.

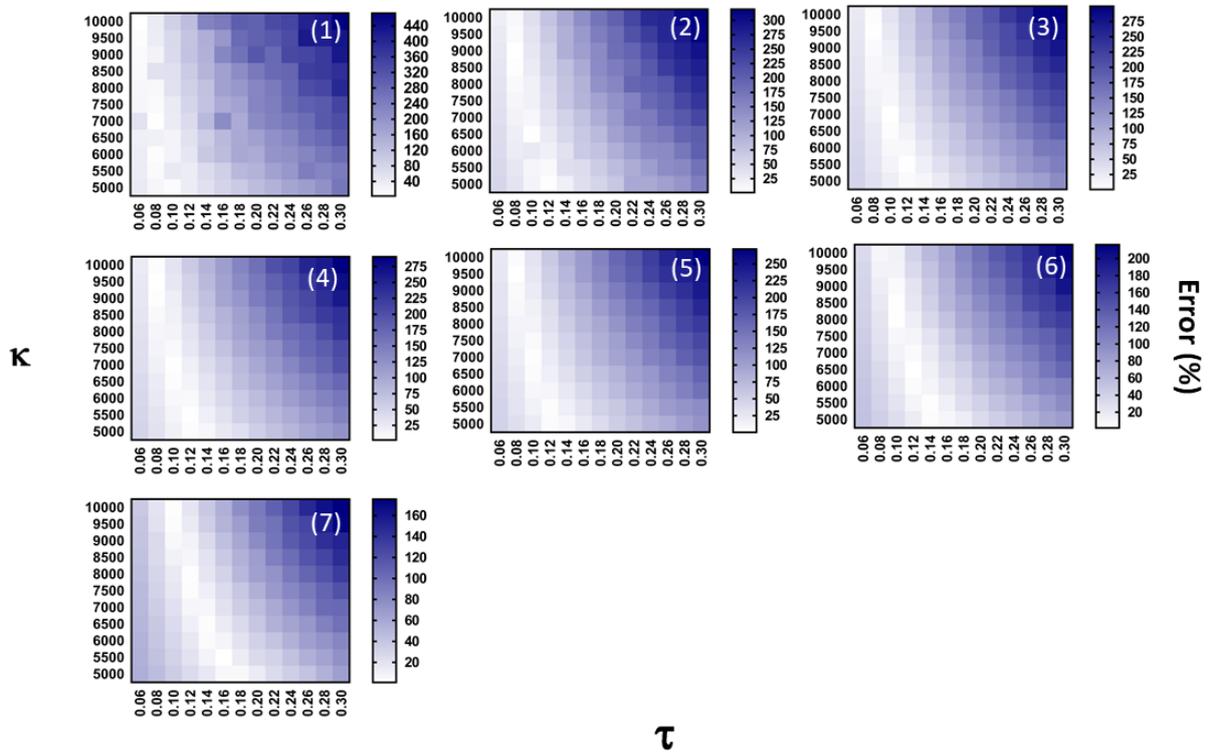

**Figure 6** A visualisation of the parameter calibration process for the key parameters of the social force layer of our simulation model based on observations of the total evacuation time obtained from Experiment IV. Plots (1)-(6) are respectively related to the bottleneck scenarios with exit width 60cm-120cm. Regions of each graph with brighter colour represent the parameter combinations that produced the least amount of simulation error for that scenario. Note that, the scale of the plot legend does not need to match across different exit widths, because for each exit width, the best parameter combinations are chosen independent of the other exit widths. The chosen parameter combination resides at the interest of the 'best parameter combinations list' for all seven different widths.

## 4. Numerical simulations setup

The numerical tests were performed on eight different simulated setups. The purpose of the numerical tests was to put the sensitivity of the simulated evacuation time estimates (as the primary metric of the evacuation simulation) under test for various parameters of the model on a comparable scale. In undertaking these numerical tests, the main factors to be considered were as follows. (1) Each simulated setup had to be complex enough to require all main layers of the model (as outlined earlier) to be active (i.e. the agent needs to make all kinds of evacuation decisions embodied by the model), (2) the sensitivity analysis to parameters at all layers of the model needed to be made on a comparable scale, (3) the analyses were conducted only for the parameters that had been directly calibrated using experimental data, and (4) the default signs of the parameters were supposed to remain intact. For certain parameters (like those of the social force model) the sign can inherently and theoretically not change, and for others (like those of the exit choice model) changing the sign of the parameter would produce behavioural patterns against what have been established through empirical testing. Our intention was to investigate the relative sensitivity of the estimates to each parameter compared to those of others while all parameters act at the direction of influence that have already been theoretically or empirically established for them.

In order to maintain the abovementioned criteria, the numerical tests were designed as follows. For each given simulated setup, and for each calibration parameter, we varied the value of the parameter by decreasing and increasing it by a factor of up to 90% of the calibrated value of that parameter. In other words, for any given parameter, assuming that the calibrated value of that parameter is $\psi$, we examine all values like $\eta \times \psi$ where $\eta$ varies between 0.1 to 1.9 at increments of 0.05 for $\eta$. In total, 13 parameters were subject of this analysis, and for each parameter, 37 different values were examined. For each of these single values, 100 simulation runs were repeated. The total evacuation time and mean evacuation time of the agents were averaged over those 100 repetitions of simulation. Therefore, the analysis required 8(setups)×13(parameters)×37(values)×100(repetitions)=384,800 repetitions in total. We did not directly record the computation times. The computation time varies across the setups depending on the complexity of the geometry, the number of simulated agents for that setup and even the value of the parameters. But for a rough average of 60secs of computation time for each run of the simulation, it is estimated that the analyses in total have taken more than 6,000 hours of computation. The computation load was distributed over ten computers and were completed over a course of four months.

Figures 7 and 8 provide a snapshot from the visualisation of all eight simulated setups (from here on, referred to as setups 1-8) as the computation is in progress. All setups were meant to simulate crowded spaces (spaces that are not heavily occupied are not subject of this analysis). Each setup needed to constitute a series of interconnected rooms with most rooms offering multiple intermediate exits to other rooms or to safety. At the onset of the evacuation, occupants exist in most spaces distributed at random positions in each room. The setups vary in terms of the shape of their geometry, the number of agents, the number of final exits (shown in red) and intermediate exits (shown in blue) and the dominant pattern of the movement in that space. Table 4 summarises some of the characteristics of these setups. For certain setups, the path to final exits required simulated occupants of multiple rooms to merge with each other and share certain spaces in order to reach the final exit(s). For those setups, we say that the simulated setup has a dominant merging pattern. For other setups the dominant pattern is that occupants scatter and distribute through peripheral exits (i.e. diverge) in order to reach the final gates. These combinations were devised to create an adequate level of variability across the simulated setups in order to make sure that the findings are not exclusive to one particular setup or geometry. In the online Supplementary Material, video samples of the simulation process for each of these eight setups (at the default calibration values) have been provided. Also, the trajectories, heatmaps of average velocities and heatmaps of average and maximum densities provided in the Appendix provide more details on the pattern of movement that each simulated setup created.

**Table 4** The list of the numerical simulation setups

| Setup No. | No. of rooms | No. of agents | No. of intermediate exits | No. of final exits | Dominant pattern of movement |
|---|---|---|---|---|---|
| 1 | 7 | 850 | 11 | 6 | Merge+Diverge |
| 2 | 5 | 1000 | 7 | 8 | Diverge |
| 3 | 9 | 450 | 12 | 2 | Merge |
| 4 | 8 | 850 | 11 | 6 | Diverge |
| 5 | 5 | 1100 | 4 | 4 | Merge |
| 6 | 5 | 1100 | 4 | 6 | Diverge |
| 7 | 4 | 700 | 3 | 3 | Merge |
| 8 | 4 | 900 | 3 | 9 | Diverge |

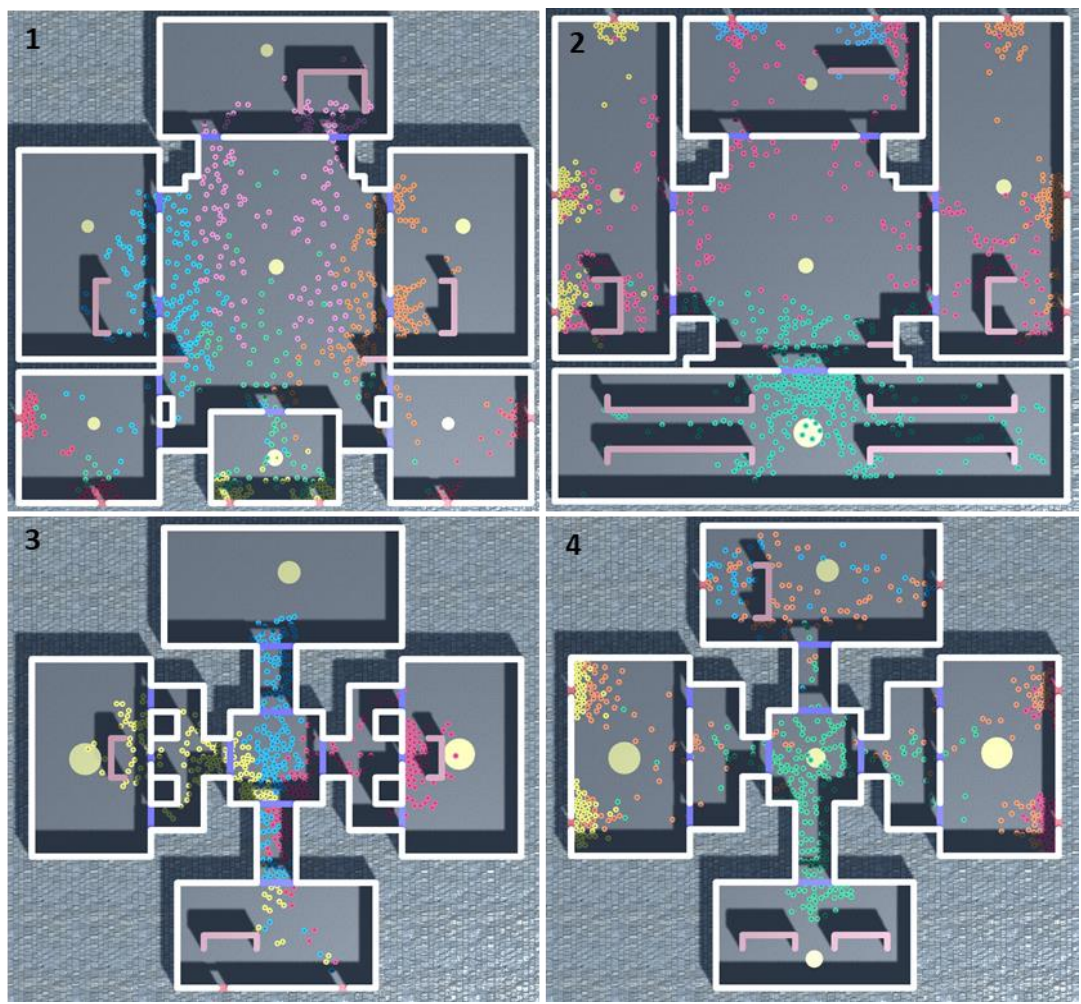

**Figure 7** Still images from the visualisation of the numerical calculation processes for setups 1-4

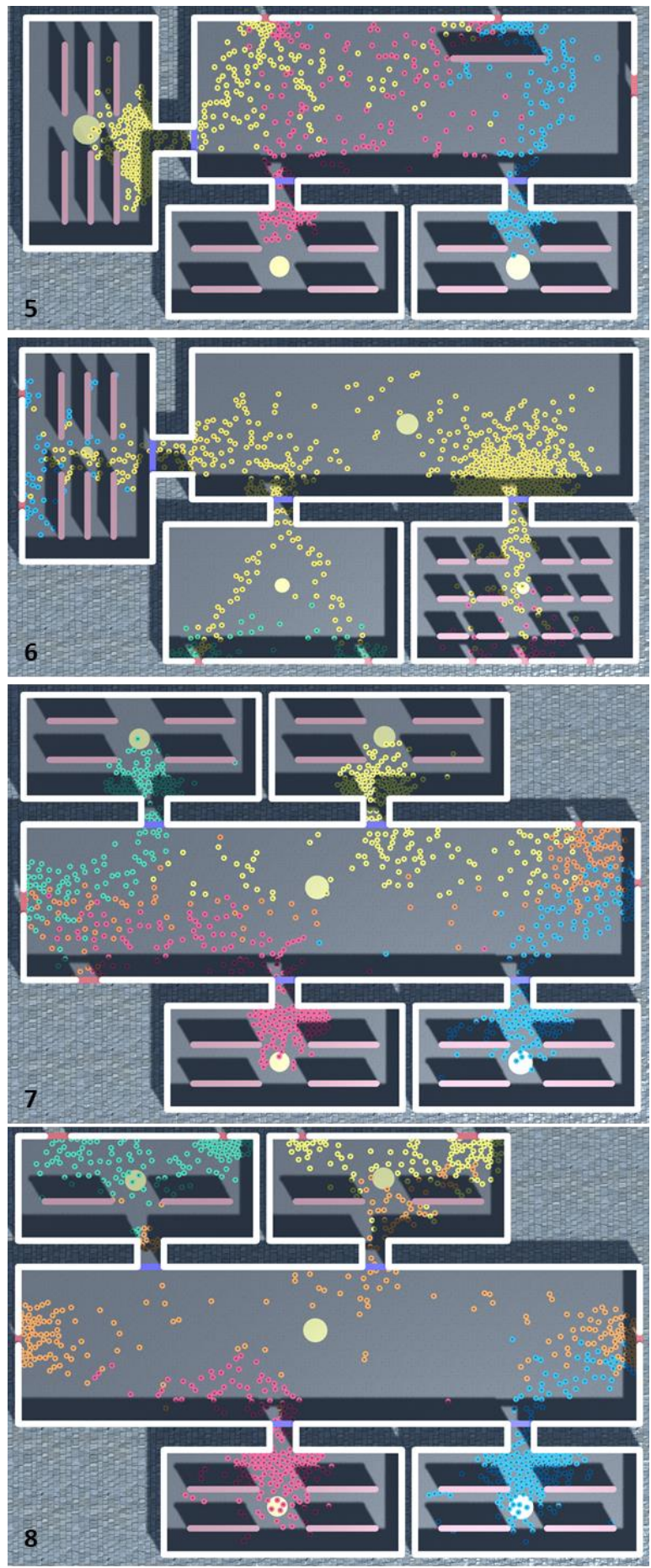

**Figure 8** Still images from the visualisation of the numerical calculation processes for setups 5-8

# 5. Computational testing results

## 5.1. Identifying the most critical parameters

The primary results of the numerical testings using the computer simulated setups and the modelling platform described earlier are illustrated in Figures 9-16 (for simulated Setups 1-8 respectively). In each set of graphs (associated with one simulated setup) we have plotted the measurement outcomes for each parameter separately. In total, 13 different parameters were the subject of these numerical testing, therefore, each graph set consists of 13 different plots. Each plot visualises the average of the total (simulated) evacuation time (denoted as Avg TET) as well as the average of the average individual evacuation time (denoted as Avg IEV) associated with each parameter value on the left and right vertical axes and using solid blue lines and red dashed lines respectively. The error band for each line-plot represents the standard deviation of the measurement. The horizontal access represents the value of the parameter through the scaling factor, $\eta$. At $\eta=1$, the simulation is performed with the calibrated value of the parameter. Similarly, $\eta>1$ means that the absolute value of the parameter is magnified relative to its calibrated value, and $\eta<1$ indicates that the parameter value it reduced in its absolute value compared to the calibrated value. The value of $\eta$ itself is stretched from 0.1 to 1.9 for each parameter. The vertical dashed line that is superimposed on each plot indicates the calibrated value of the parameter represented by that plot (i.e. $\eta=1$). While the horizontal axes of all the plots are made consistent through the scaling factor, $\eta$ that is common across them, the vertical axes are also made consistent in order for the sensitivities to be fully comparable across different parameters. The range of the vertical axis's values associated with each graph set (i.e. associated with each simulated setup) is determined by the parameter to which the simulated measurements showed the highest degree of sensitivity. As a result, the range of the vertical axes is consistent for the plots associated with each simulated setup but they differ from graph set to graph set (i.e. they differ across the simulated setups).

While the numerical results of these simulation testings showed certain degrees of case specificity, one aspect remained consistent across various setups. The outcome of the simulation is by far margin more sensitive to the two mechanical movement parameters, $\kappa$ and $\tau$, compared to the parameters related to the decision-making aspects. This is shown by the fact that, once we made the range of the vertical axes consistent across different plots for each setup, the sensitivity to most decision-making parameters paled into insignificance compared to the sensitivity to $\kappa$ and $\tau$ values. As a result of this, the majority of the line plots appeared horizontal for these decision-making parameters. The level of sensitivity to these mechanical parameters were highly substantial. And between the two of them, the value of $\kappa$ was even more critical. For example, according to the outcomes of the simulation analyses for Setup 1, the simulated total evacuation time may vary between 30 seconds and 105 seconds through stretching the value of $\kappa$ by a factor of 90% to each direction (from its calibrated value). And these figures for Setup 2 are respectively 40 secs and 160 secs. The pattern of the variation of the simulation evacuation time in response to changing the value of $\tau$ was by and large linear across all eight setups but the pattern was more of a piecewise linear nature for $\kappa$, with the gradient of the change suddenly increasing for values greater than $\eta \approx 1.5$. Also, the pattern of variation was almost invariably consistent across our two measurements, total and average of individual evacuation times. This indicated that for analyses such as that of this work, as well as for evacuation optimisation programming ([Abdelghany et al., 2014](#)) either of these two measures may be used interchangeably.

The results of this part of the analyses indicated that, for complex, geometric setups that involve interconnected spaces and relatively large crowds compared to the exit capacities of the space (i.e.

where capacity is restrictive and bottlenecks ought to be created as a result), the sensitivity of the simulated measures of evacuation to the mechanical movement parameters (in particular, those that determine the flowrate at bottlenecks) outweigh the sensitivity to decision-making parameters by a far margin. This means that for setups where the capacity of the space is such that bottlenecks are created, the simulated measures of total (or average individual) evacuation times are by far degrees dependant to how we specify the value of these mechanical parameters (those that determine the level of physical friction between individuals or the intensity of the movement drive of each individual). Altering the value of these parameters makes a great difference in our predictions in a way that altering the value of the decision-making parameters in most instances does not compare with it. Changing the value of these mechanical parameters by a factor of 50% of its original value, for example, can change the outcome of the simulated evacuation time by a factor of nearly 20% (according to Setup 1, as an example). Whereas making the same relative magnitude of alteration to the value of the decision-making parameters does not make such large difference in the simulated evacuation time. This does not, by any means, indicate that the patterns of movement in the simulation process will not change as the decision-making parameters are altered. Certain parameter specifications for these decision parameters produce less plausible and less accurate movement patterns, and their value do influence the outcome of the simulation. But this change of the collective movement pattern does not reflect as much in the evacuation time estimates as the change caused by the bottleneck-flowrate-related parameters. The practical implication of this observation is that a certain percentage of error in calibrating the value of these mechanical parameters biases our predictions at a much greater magnitude compared to the same percentage of error in calibrating decision-making parameters. Besides, it should be noted that the impact mis-specification of these mechanical parameters are hard to be detected visually, whereas for decision-making parameters, poor parameter setting will create implausible movement patterns that can often be visually observed.

It is also important to note that here, we are talking about the unilateral change of a single parameter value while other parameters are still set at their calibrated value. Therefore, the analyses are not meant to downplay or trivialise the significance of calibrating the decision-making parameters. Their importance will be more appreciated if did not have access to those based values calibrated directly through experiments and needed to set all these unspecified values simultaneously. The fact that all parameters other than the one that is subject of the sensitivity analysis are set at their appropriate values is a major factor in not allowing the outcomes to drastically change. In other words, a fully calibrated model may make single individual parameters to become robust to mis-estimation (or calibration error). For example, when a model includes a decision-changing module whose parameters are well calibrated, the model may become more robust to the mis-specification of the individual exit-choice parameters. In other words, even when the value of an individual exit-choice parameter is altered from its calibrated value, the mere fact that the agents are allowed to change their exit choices during their movement will alleviate the effect of mis-specifying the value of that single parameter at the exit-choice layer. Furthermore, other parameters of the exit-choice module will still be at their calibrated value and that too alleviates the unilateral effect of that single mis-specified exit-choice parameter. But the two mechanical movement parameters in our analyses, in comparison, have a more direct impact on the movement, and that effect is not necessarily compensated by other layers or parameters of the model. Also, here we are not considering cases of 'gross parameter mis-estimation'. Rather, the variation to the value of each individual parameter is within a range of 10% to 190% around its calibrated value. We did not consider cases where the sign of the parameter is specified incorrectly or the cases where the magnitude of the parameter is grossly mis-specified.

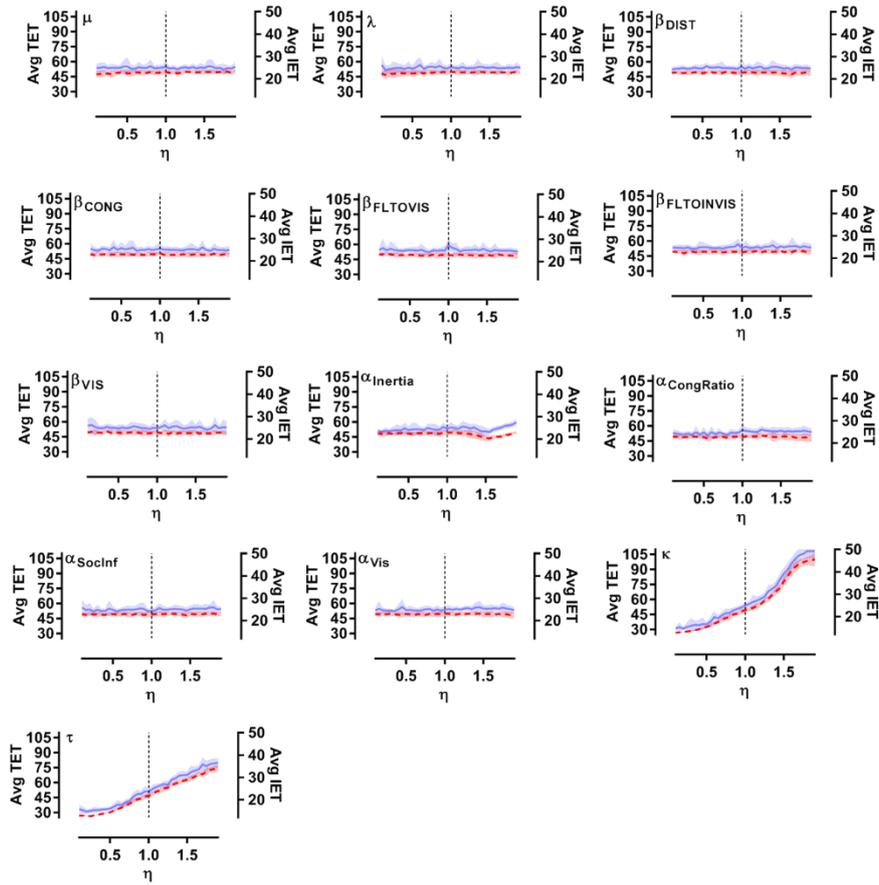

**Figure 9** Outcomes of the numerical analysis for the simulation setup 1

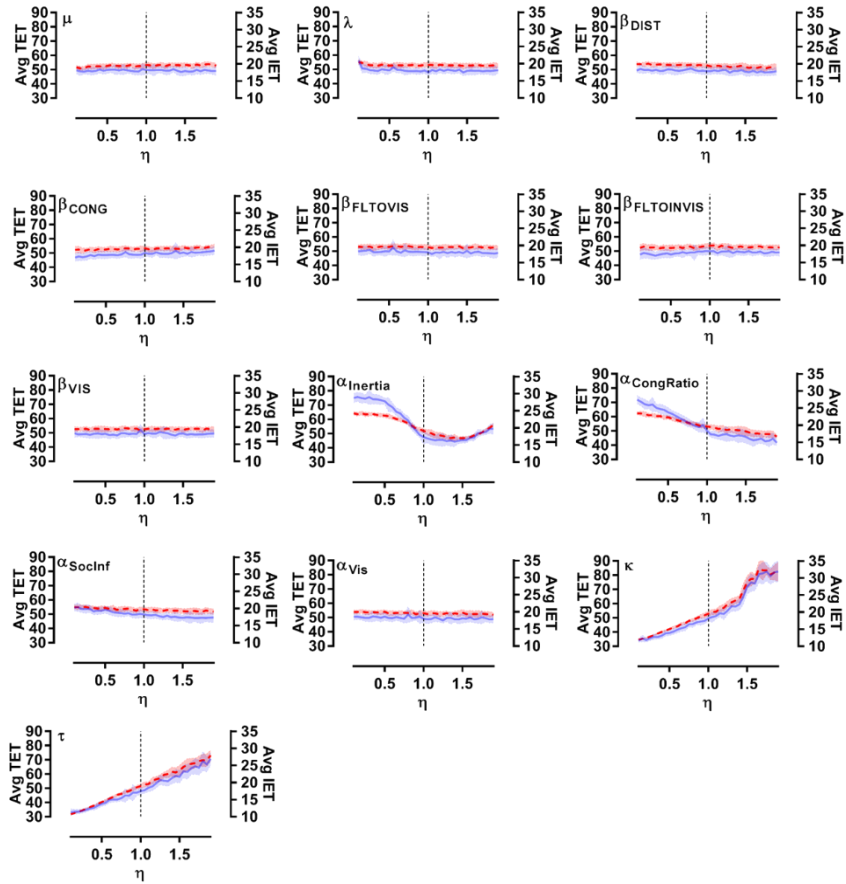

**Figure 10** Outcomes of the numerical analysis for the simulation setup 2

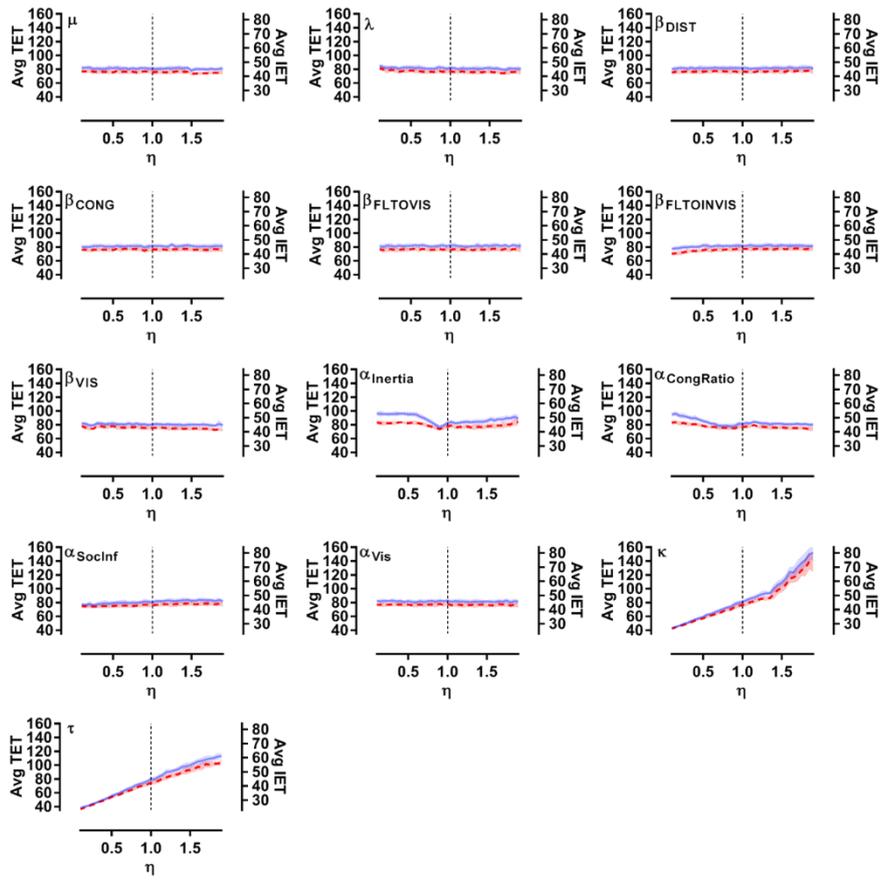

**Figure 11** Outcomes of the numerical analysis for the simulation setup 3

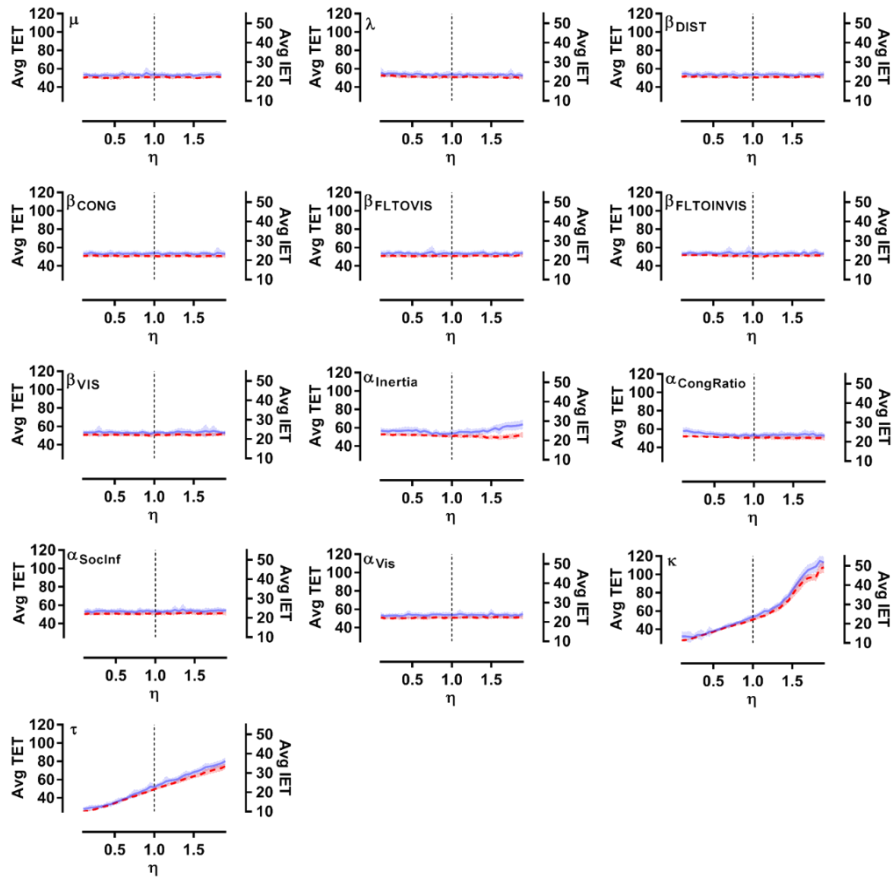

**Figure 12** Outcomes of the numerical analysis for the simulation setup 4

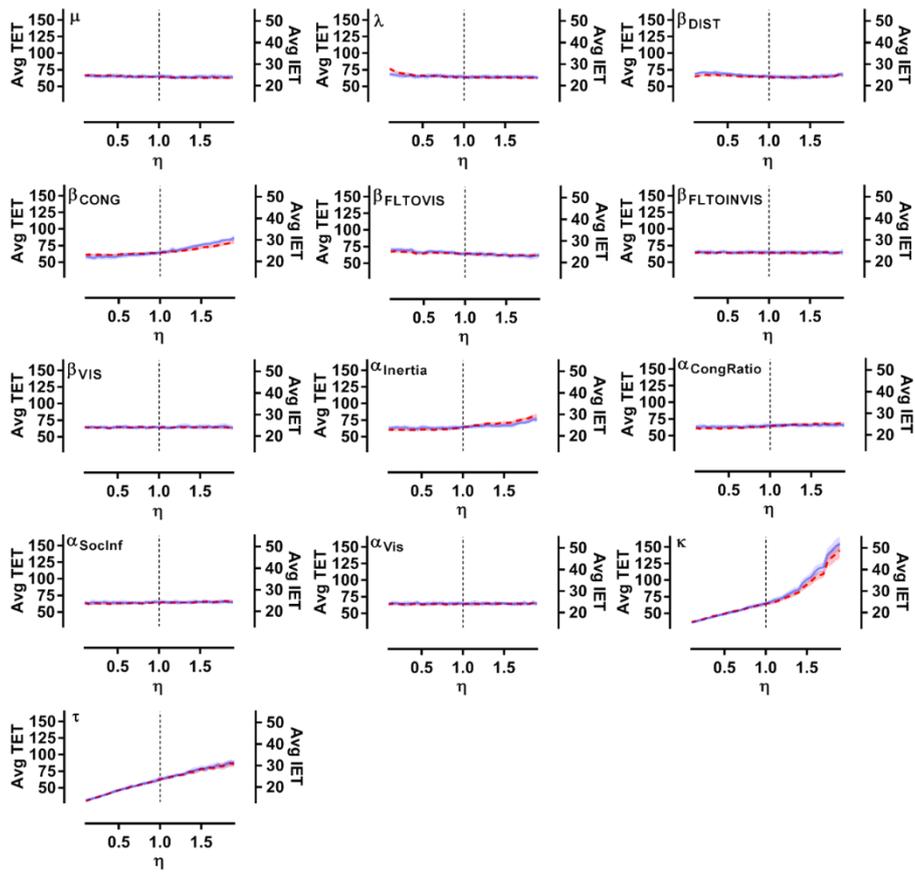

**Figure 13** Outcomes of the numerical analysis for the simulation setup 5

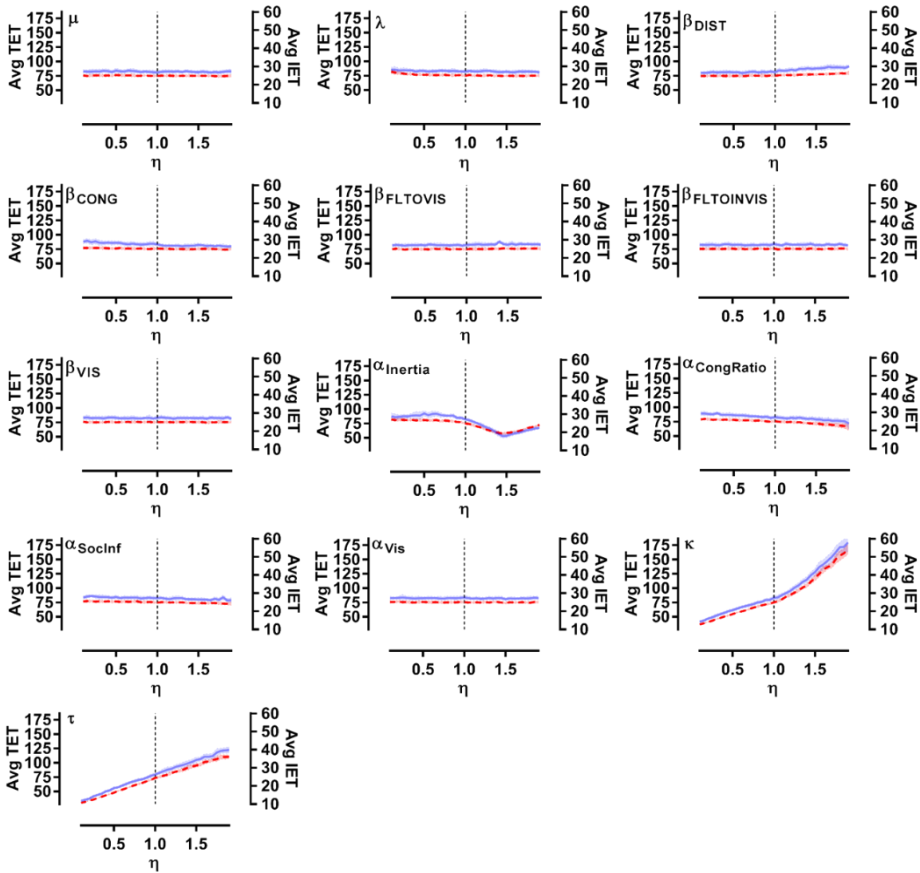

**Figure 14** Outcomes of the numerical analysis for the simulation setup 6

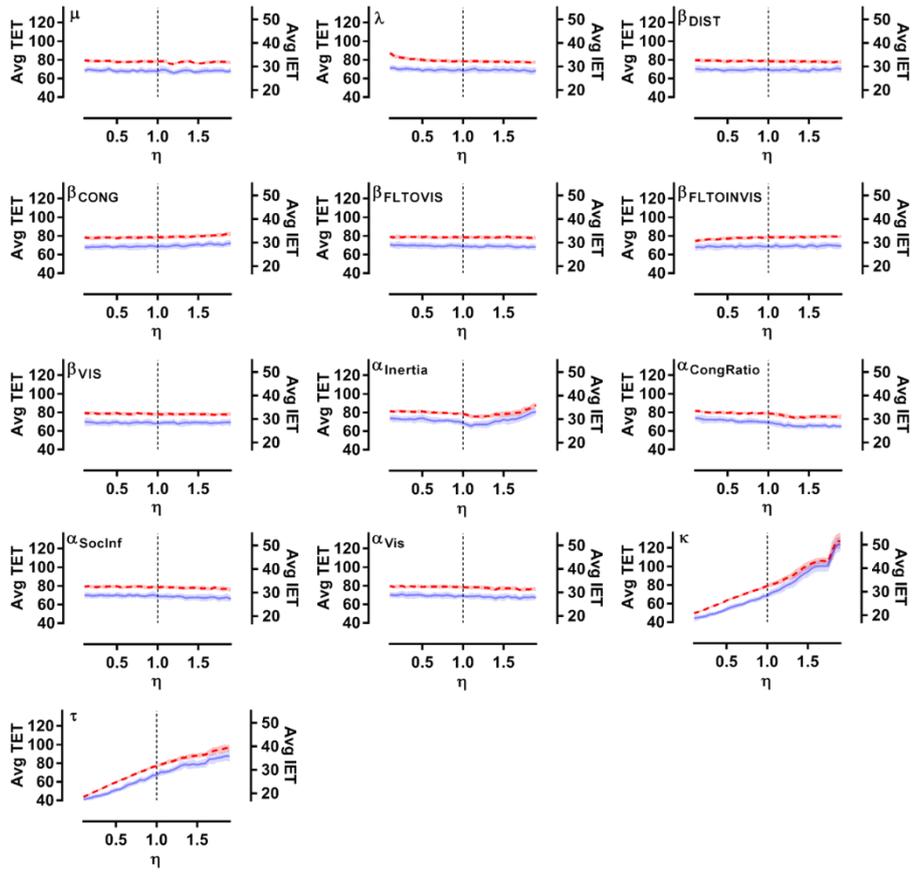

**Figure 15** Outcomes of the numerical analysis for the simulation setup 7

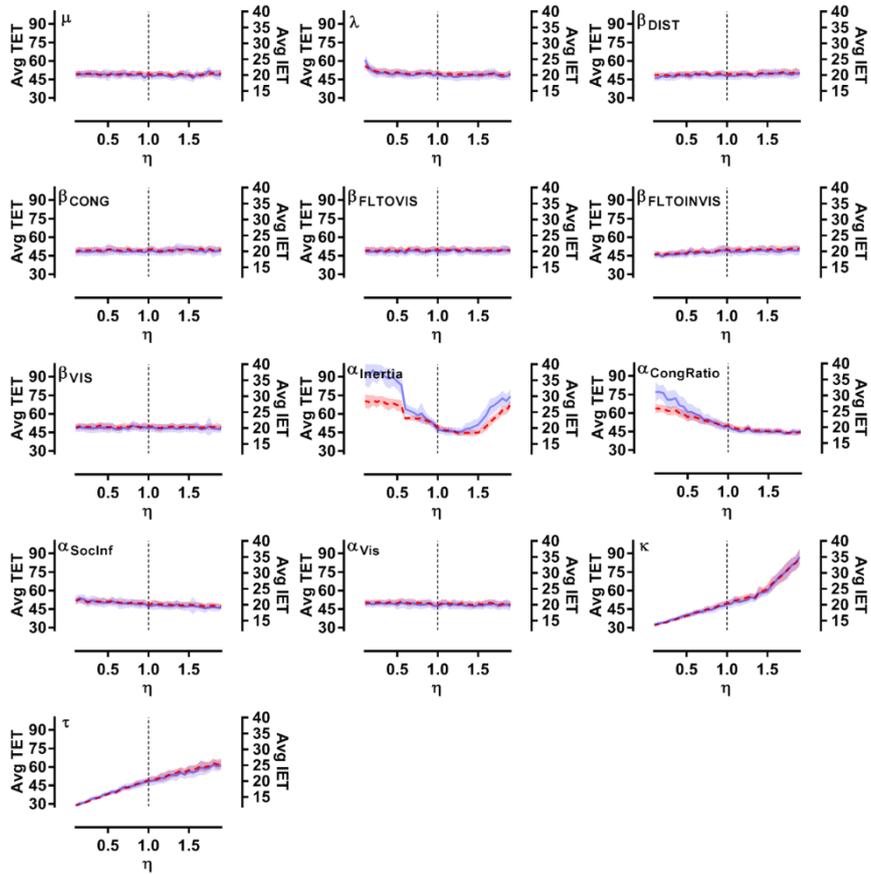

**Figure 16** Outcomes of the numerical analysis for the simulation setup 8

*5.2. Inferring behavioural insight*

Apart from the main purpose of this analysis, which was comparing the relative sensitivity of the aggregate simulated measures to the value of individual parameters at various modelling layers, there are also secondary outcomes offering behavioural insight to be drawn from such analyses. This is because of the fact that almost all individual parameters at the decision-making layers of the model employed in this work have certain behavioural interpretation. As a result, altering the value of those parameters is equivalent of altering certain aspects of evacuee's behaviour in certain directions. The impact of these behavioural modifications, however, was mostly masked out by the fact that we set the scale of the vertical axes in the same way across different parameters for each simulated setup. In order to obtain such type of insight from these numerical testings, we have singled out from each set of plots those other than the ones representing $\kappa$ and $\tau$ that showed the greatest degree of sensitivity. We then replotted them by re-adjusting their vertical axis scale and customising it for each plot in a way to effectively observe the pattern of variation within that graph.

These singled out plots are illustrated collectively in Figure 17 (for simulation setups 1-5) and Figure 18 (for simulation setups 6-8). The majority of the plots that made to this list (17 out of 25) are related to parameters related to exit-choice-changing layer of the model. Next is the parameters related to the exit-choice layer of the model (6 out of 25). Also, in 2 instances (2 setups), the results showed considerable sensitivity to the value of the reaction time parameters of the model. Also, the majority of the exit-choice and exit-choice-changing parameters that made to this list were related to the inter-individual interactions or the so-called social influence parameters (such as CONG from the exit-choice layer and CongRatio and SocInf from the exit-choice-changing layer). An exception to this is the Inertia parameter which is not related to social interactions but has appeared 8 times in this list, more than any other parameter.

*Sensitivity to the Inertia parameter.* The inertia parameter is directly correlated with the frequency of the exit-choice changes without regard to the contextual variables at the moment of simulated time when the decision change possibility is offered to the simulated agent. Greater absolute values for this parameter means that the agent is more resistant towards changing the initial exit decision. Smaller values, on the other hand, generate greater likelihood for the agent to change its initial exit choice. The estimation of the simulated evacuation time showed substantial sensitivity to the value of this parameter in all 8 setups that we examined. This indicates that the frequency and likelihood of decision changing is a major factor that determines (1) the prediction of the simulated evacuation time in computational settings and (2) the actual evacuation time in real-world scenarios. The pattern of the variation of the evacuation time with respect to the value of this parameter showed very little degrees of case specificity. With the exception of Setup 5 where the simulated evacuation time varies rather monotonically with respect to the scale factor $\eta$, in other cases there is a minimum in the plot at the intermediate values of the scale factor. This indicates that during an evacuation process, certain intermediate degrees of 'openness' to revising exit choices by individuals will collectively result in better outcomes (i.e. shorter evacuation times) for the system. It appears that neither of the two strategies of 'too much inertia' or 'or too little inertia' are the optimum level of tendency in regard to the exit-choice changing behaviour. Rather, an intermediate degree of inertia is often most optimum for the system. Also interestingly, this minimum point often occurred in our analyses at scale values of $\eta>1$. Considering that $\eta=1$ is associated with the calibrated value for the inertia parameter reflecting the natural (observed) tendency of people, this indicates that systems of crowd evacuation could potentially become more efficient by having people to increase their openness or willingness to be more actively adaptive in terms of their exit choices. This could be one of the ways through which one can make improvements in evacuation efficiency through simple behavioural modifications.

*Sensitivity to the CongRatio parameter.* The CongRatio parameter of the decision-changing model also correlates with the frequency of the exit-choice changes by relating the likelihood of change to the extent of the queue-size imbalance at exits. Greater absolute values for this parameter mean that the simulated agent is more sensitive to the presence and extent of the congestion imbalance, and is thus, more likely to make a decision change when such imbalance exists (i.e. when the previously chosen exit is performing relatively poorly in terms of the queue size compared to the least congested exit in that room). Smaller absolute values mean that the agent take this factor less importantly in choosing between the change and status-quo options. The estimations of the simulated evacuation times in these analyses revealed that the system is considerably sensitive to the value of this parameter. In 6 out of the 8 cases that we examined, the sensitivity was readily noticeable. With little variability across these setups, the simulation outcomes suggest that greater values for this parameter results in shorter evacuation times. This indicates that when individual evacuees show greater sensitivity to the relative performance of their previously-chosen exit, the process of evacuation becomes more efficient and that pattern is rathe monotonic, meaning that this greater sensitivity is almost invariably beneficial to the system.

*Sensitivity to the SocInf parameter.* The SocInf parameter of the decision-changing model correlates with the frequency of the exit-choice changes by relating the probability of change to whether or not other people have made decision changes from that exit moments earlier. Greater absolute values for this parameter mean that the simulated evacuee is more sensitive to observing other people making decision changes and is more likely to imitate those who make decision change right before him/her. Similarly, smaller absolute values mean that neighbours' behaviour has little effect on the decision changing probability. In relative terms, the impact of this parameter's value on the system was less noticeable, at least compared to the Inertia and CongRatio parameters. But in certain setups, like Setup 3, 7 and 8, very moderate degrees of system sensitivity were observed. The pattern of variation of the evacuation time was, however, case specific. In Setup 3, greater values mean longer evacuation times meaning that imitation in exit-choice changing was detrimental. But in Setups 7 and 8, the pattern of variation was decreasing (though at a non-steep slope) meaning that imitation in exit-choice changing was beneficial. However, it is important for us to emphasise that in terms of the magnitude of the impact, this parameter was not so influential on the system performance.

*Sensitivity to the DIST parameter.* In two cases, Setups 5 and 6, we observed rather noticeable degree of evacuation time sensitivity to the DIST parameter of the exit-choice model. Greater absolute values of this parameter mean that the simulated evacuee has a greater preference for choosing exits that are closer, and smaller absolute values mean that the evacuee does not have a string preference for choosing the nearest exit. The variation of the evacuation time to the value of this parameter in Setup 5 showed a non-monotonic pattern with a minimum occurring at the intermediate values. Whereas for Setup 6, evacuation time monotonically increased as the absolute value of the DIST parameter increased, meaning that greater preference for choosing the nearest exit was detrimental to the overall evacuation process.

*Sensitivity to the CONG parameter.* The sensitivity of simulated evacuation times appeared to be noticeable to the value of the CONG parameter of the exit-choice model in three cases, Setups 5, 6 and 7. The value of this parameter determines the degree of relative weight that the evacuee gives to exit congestion factor when making exit choices. Greater absolute values of the parameter mean that the evacuee places a high level of priority on choosing less congested exits and smaller values indicate that the congestion becomes of less importance in exit-choice trade-offs. In Setup 5 and 7, greater values of the CONG parameter resulted in longer evacuation times, whereas in Setup 6, the opposite was the case.

*Sensitivity to the λ parameter.* The λ parameter of the reaction-time model is directly related to both mean and variance of the simulated reaction times, with smaller values resulting in greater mean and variance. In other words, when λ is set at smaller values, this results in greater average delay in movement initiation as well as greater variability in the movement initiation moment across evacuees. On the other hand, greater λ values result in more instant movement reactions by the population of simulated evacuees. In two cases, Setups 5 and 7, we observed rather noticeable dependency of the evacuation time on the value of this parameter, and in both cases, greater λ values resulted in shorter evacuation times. This indicates that in those cases, the system became more efficient when simulated evacuees showed a more instant reaction rather than delaying their movement initiation.

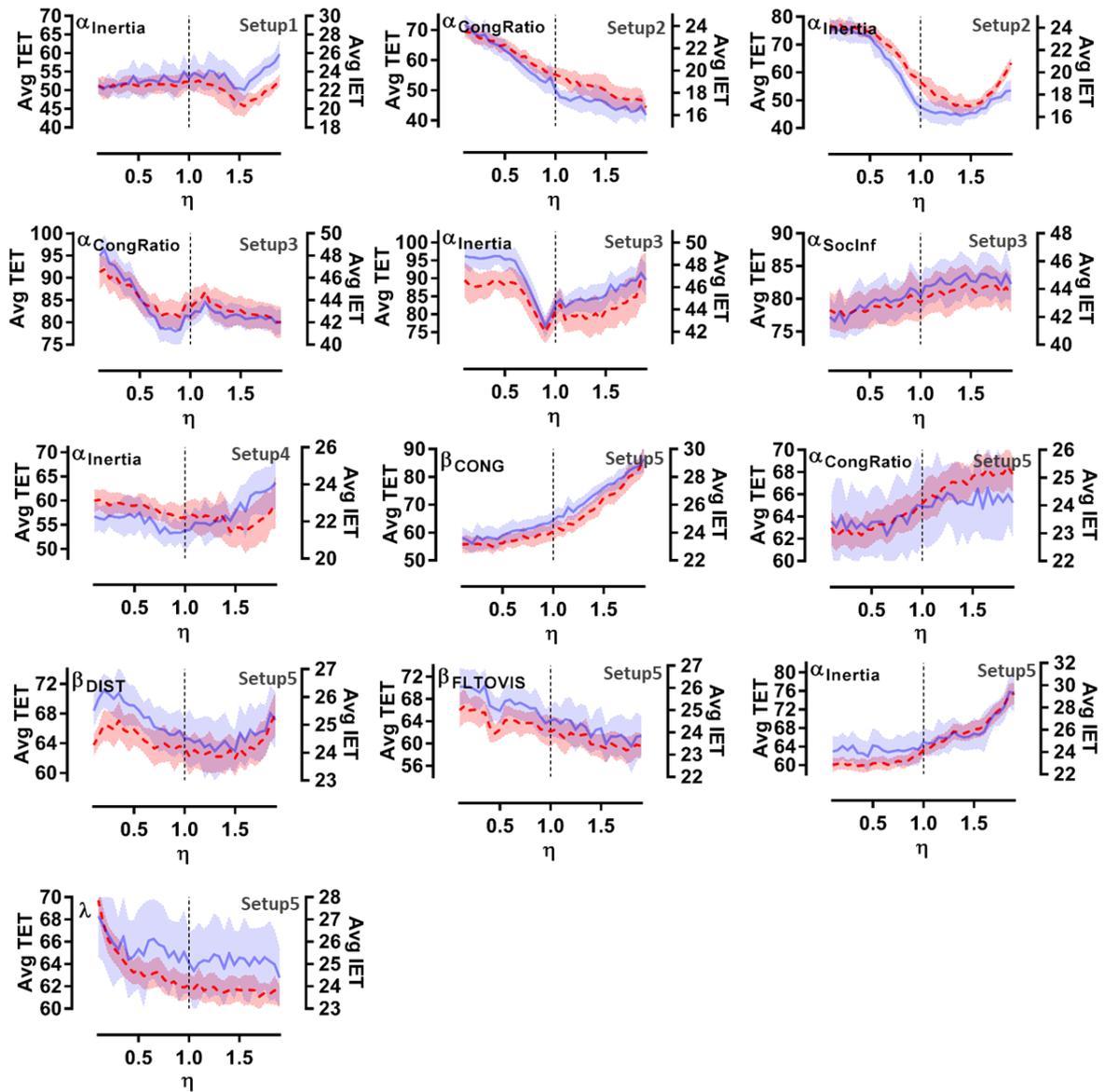

**Figure 17** Outcomes of the numerical analysis for a selected subset of parameters (except for the social force parameters) to which the simulated evacuation times showed most amount of sensitivity in setups 1-5. The scale of the vertical axis has been adjusted and customised for each graph in order to better display the degree of variation.

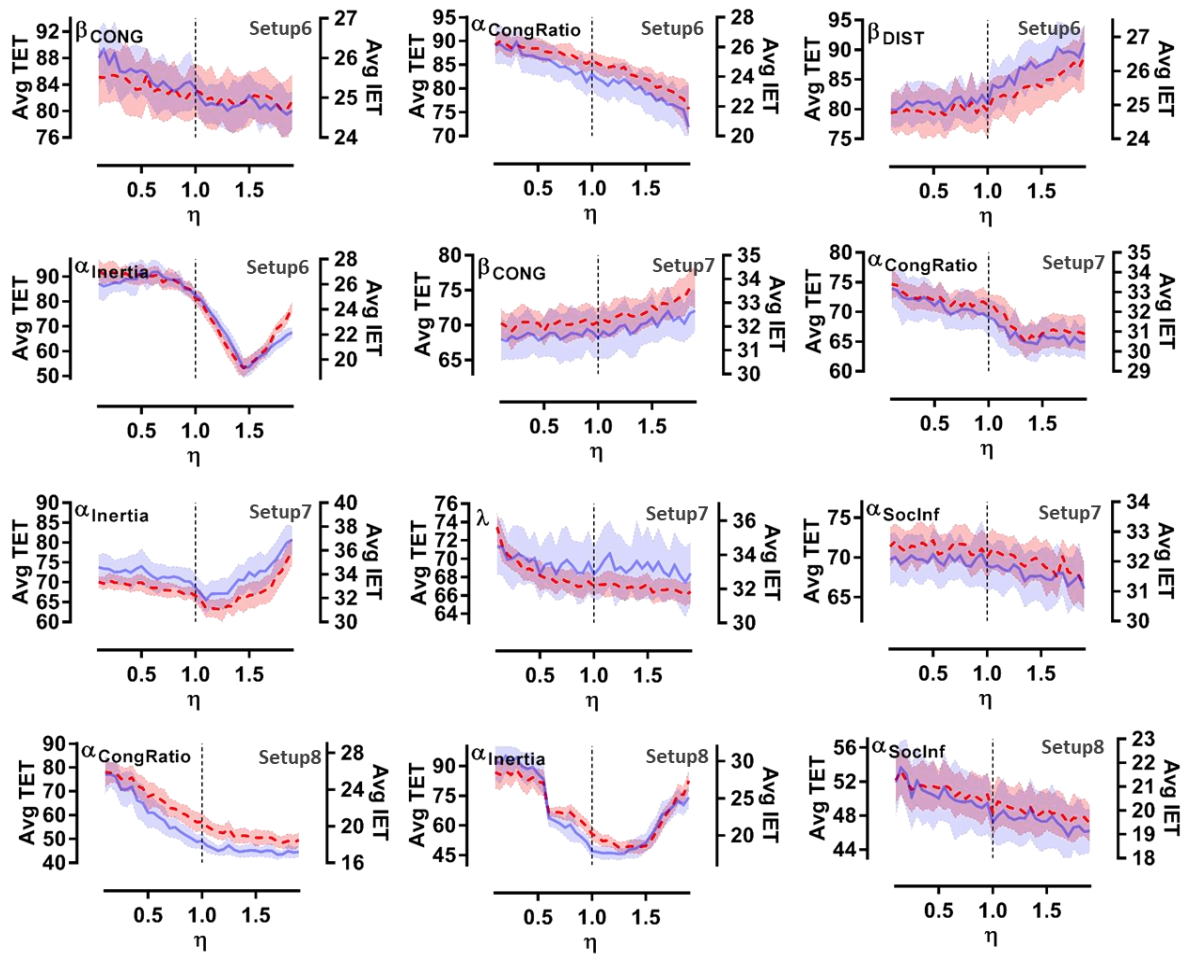

**Figure 18** Outcomes of the numerical analysis for a selected subset of parameters (except for the social force parameters) to which the simulated evacuation times showed most amount of sensitivity in setups 6-8. The scale of the vertical axis has been adjusted and customised for each graph in order to better display the degree of variation.

# 6. Conclusions

Simulation of a crowd evacuation process requires the analyst to specify the value of a range of measurable and non-measurable inputs. The non-measurable inputs often include the calibration parameters. In agent-based micro-simulation models, the analyst needs to deal with a variety of these parameters and fine-tune their values in order to produce realistic outcomes. These parameters reflect various aspects of evacuee's behaviour and have important implications for the plausibility of the predictions and simulated estimates.

This study was a first attempt to quantify the relative impact of specifying the value of these various parameters on major simulation metrics, such as simulated evacuation times. In doing this comparative analysis, we employed a multi-layer micro-simulation model whose decision-making layers are predominantly based on econometric choice and duration models. First, we provided an overall overview of how these various layers of the model can directly and separately be calibrated using individual-level observations collected from experimental settings. Having set these empirically calibrated parameter values as the benchmark and by introducing a scaling factor to magnify and decrease the value of various parameters on a consistent and comparable scale, we investigated the abovementioned question through extensive numerical testings.

Our numerical testings considered scenarios in which the simulated space is relatively complex in a way that it requires all kinds of decision-making layers to be active in the model. They also exclusively focused on scenarios where the occupancy level is high relative to the exit capacities in a wat that it creates bottlenecks and relatively complex decision trade-offs during the process of evacuation. Our computational analyses revealed that, in such scenarios, the parameters of the mechanical movement layer of the model that determine the flowrates at bottlenecks are, by far, most impactful in terms of determining the total evacuation times. A small percentage of change to the value of those parameters (relative to their calibrated values) invariably caused considerable changes to the prediction of the evacuation time. The magnitude of the impact was comparatively far greater than the magnitude of the impact resulted from similar percentage of change in the value of the decision-making parameters. Followed by these parameters, were the parameters that influence the exit-choice-changing behaviour of the simulated evacuees.

Simulation of an evacuation is always subject to prediction and modelling error and every computational model will be to some degrees vulnerable to errors of various sources including those resulted from parameter mis-specification. These observations, however, indicated that simulation of crowded scenarios is not equally sensitive to the calibration of the parameters at all layers. With certain parameters, a given percentage of error in calibration engenders greater error in terms of the simulated evacuation time compared to the same percentage of error in calibration of other parameters. Therefore, if resources are limited for experimentation or data acquisition, those parameters should reasonably be prioritised. Also, these observations warrant that the analyst devise more rigorous methods, collect better quality data or increase the sample size and perform further validation testing specifically for those particular parameters.

In simple terms, it appears that in order to predict evacuation times of crowded scenarios accurately, the most important aspect is getting the bottleneck flow rates right. That means that (i) we need further empirical testing and data collection on bottleneck flows (Haghani et al., 2019c; Liao et al., 2014; Nicolas et al., 2017; Seyfried, 2009; Seyfried et al., 2010; Tian et al., 2012), (ii) we need to increase the accuracy of our empirical estimates for bottleneck flow rates, (Daamen and Hoogendoorn, 2012b; Tobias et al., 2006) and (iii) we need more rigorous methods for calibration of the bottleneck parameters (Daamen and Hoogendoorn, 2012a). Otherwise, our efforts in making more sophisticated decision-making models or calibrating their parameters may be largely in vain.

The literature is making substantial progress in improving our empirical knowledge of bottleneck flows as one of the most extensively studied areas in this field. The emerging empirical testings are showing how that the bottleneck capacity is a functions a whole range of factors other than the width of exits, factors such as the level of jam at the bottleneck (Cepolina, 2009; Cepolina and Farina, 2010), the behavioural conduct of evacuees (Gwynne et al., 2009; Yanagisawa and Nishinari, 2007) or the physical characteristics of the exit (Daamen and Hoogendoorn, 2012b). We are observing increasing evidence suggesting that the capacity-width relationship is not linear (Gwynne et al., 2009; Hoogendoorn and Daamen, 2005). But, are our existing models nuanced enough to represent all these empirically-established phenomena accurately? In fact, in our experience, we have observed that certain sets of bottleneck parameters perform very nicely for certain exit widths but not as much so for other exit widths. The set of the bottleneck parameters that we chose in our calibration process was a best compromise between these. We chose the set of values that overall showed best performance on various exit widths. And that is because of the fact that there is a tendency to have one single set of calibration values for all circumstances and we concede that this may be one of the many factors that prevents us from achieving higher degrees of accuracy in evacuation simulation. A simple inspection of Figure 6 clearly shows that for different exit widths, different sets of $\tau$ and $\kappa$ display the best performance. Therefore, if that is the case, the question is why not making our parameters at bottlenecks exit-width-specific? That could be only one way among many areas to improve the accuracy of estimating bottleneck throughput rates.

The previous statements were not meant to trivialise the significance of calibrating decision-making parameters in simulation models. To the contrary, we have indeed observed how these parameters can influence the plausibility of the movement patterns and the simulated evacuation times. Our finding only highlights that 'in relative terms' fine-tuning the bottleneck throughputs have a larger impact and thus need to receive higher priority.

It is also interesting that among the decision-making parameters, those that comparatively had the most noticeable impact are those that have received the least amount of attention in the literature so far, the decision-changing parameters (Gwynne et al., 2000; Haghani and Sarvi, 2019a). Within the last few years, a considerable amount of experimentation, data collection and modelling effort has been dedicated to exit-choice modelling (Bode and Codling, 2013; Haghani and Sarvi, 2016, 2017; Liao et al., 2017; Lovreglio et al., 2014a), whereas very little has been comparatively done towards decision adaptation aspect of the behaviour. The analyses in this work indicate that again, without an adequately accurate model of exit-choice changing, our efforts in developing more sophisticated exit-choice models or collecting observation for that might be wasted. It is imperative from the modelling perspective that we be able to maintain a consistent level of accuracy at various levels of these models, otherwise, perfecting one or two layer(s) and leaving the others our of a rigorous calibration process might not necessarily produce the desired outcome.

Another aspect of our study was the behavioural findings that we observed through performing the sensitivity analyses testing on decision-making parameters. Since decision-making parameter each carry certain behavioural indication, changing their value is basically equivalent of changing evacuees' behaviour and tendencies in decision making. And as Gwynne and Hunt (2018) have aptly pointed out, these types of parametric agent-based models provide us with a valuable opportunity of turning the model to a "computational behavioural laboratory" and examining the impact of various behavioural strategies on evacuation efficiency. We observed, for example, how lesser inertia in exit-choice making or more sensitivity to congestion imbalance can potentially shorten total evacuation times if adopted by individual evacuees. We believe this (i.e. behavioural optimisation) could be a potentially overlooked approach in evacuation optimisation whose benefit has not yet been fully explored.

This work was limited to examining the impact of non-measurable inputs of simulation models, the behavioural parameters. However, measurable inputs (such as body size, body mass, desired velocity etc) are also factors that may potentially make great influence on simulation estimates. There needs to be further analyses on those aspects to quantify their relative significance and to improve our empirical estimates for those inputs. Also, here, we only limited our testings to a single platform of modelling. A question that is comparable to what we studied here concerns the degree to which the predictions are sensitive to different structures of modelling at various layers. For example, we used a combination of weighted shortest pathfinding algorithm and social-force model for our locomotion layer of modelling. It is not however clear to us to what degrees these simulation model outputs are sensitive to the use of alternative locomotion models such as those based on cellular automata methods (Bandini et al., 2011; Burstedde et al., 2001; Guan et al., 2016), or discrete choice walking behaviour methods (Antonini et al., 2006; Robin et al., 2009). Similarly, one may ask to what degrees the simulated outputs are sensitive to model specification at the decision-making layers. Would it make a meaningful difference, for example, if we shift our exit-choice model from utility maximisation to regret minimisation or our reaction time model from Cox-proportional Weibull to an equivalent Exponential model? Does model specification have a bigger impact on our predictions than parameter calibration or even measurable input calibration? Identifying the aspects of the modelling among all these dimensions that need to be prioritised in evacuation modelling could be a major step forward in improving our computational accuracy.

**Appendix**

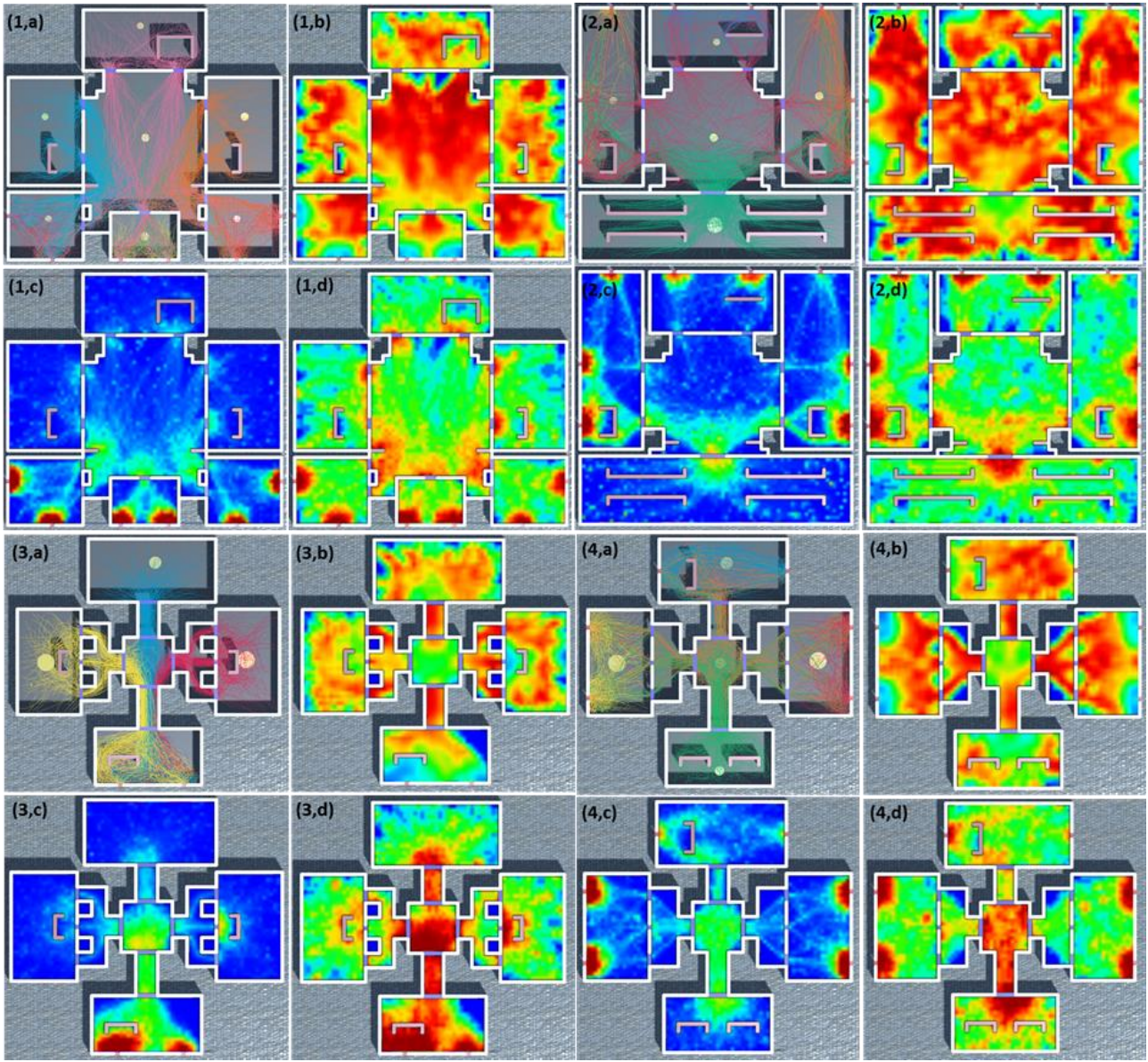

**Figure A1** Visualisation of the mass trajectories of simulated agents (subfigures (a)), maximum density (subfigures (b)), average density (subfigures (c)) and average velocity (subfigures (d)) for the simulation setups 1-4. The first component of each label signifies the number of the setup. The outputs are related to the calculations based on the default parameter setting of the numerical model.

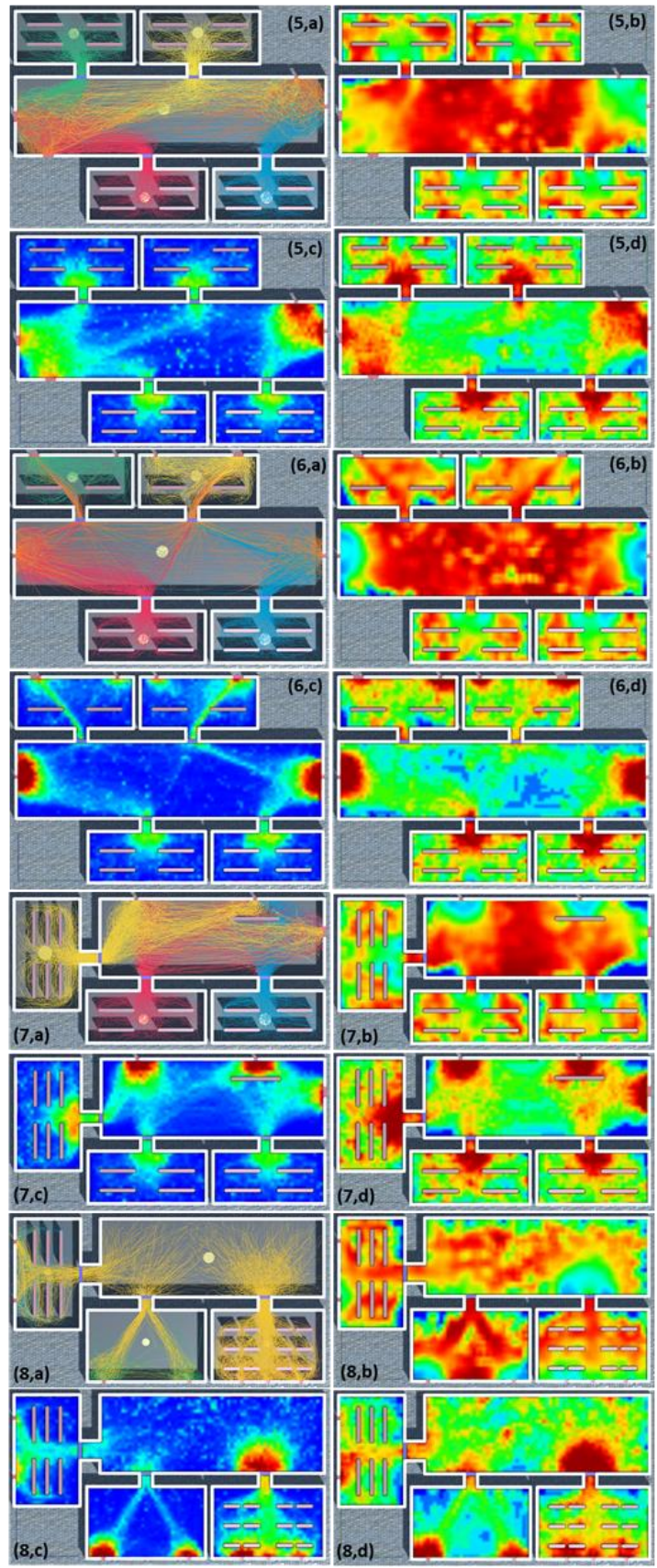

**Figure A2** Visualisation of the mass trajectories of simulated agents (subfigures (a)), maximum density (subfigures (b)), average density (subfigures (c)) and average velocity (subfigures (d)) for the simulation setups 5-8. The first component of each label signifies the number of the setup. The outputs are related to the calculations based on the default parameter setting of the numerical model.

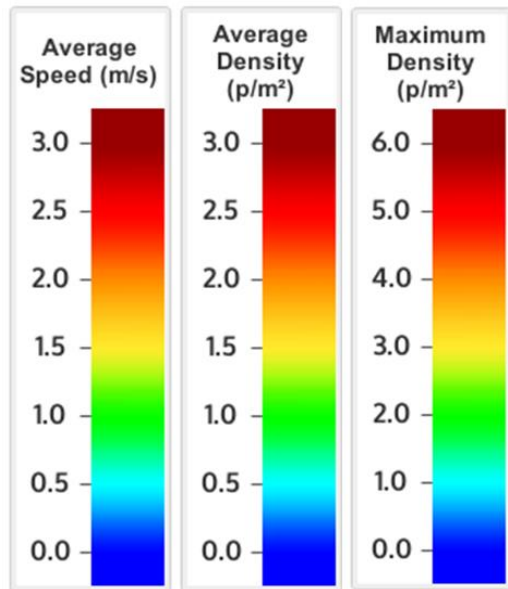

**Figure A3** Legend for the heatmap plots in Figures A2 and A3